\documentclass[12pt,peerreviewca,draftcls,onecolumn]{IEEEtran}

\usepackage{amsmath}
\usepackage{cite}
\usepackage{caption3}
\usepackage{graphicx}
\usepackage{paralist}
\usepackage{rotating}

\newcommand{\figsz}{5}

\begin{document}

\title{The Error-Pattern-Correcting Turbo Equalizer}

\author{Hakim~Alhussien,~\IEEEmembership{Member,~IEEE,}
        and~Jaekyun~Moon,~\IEEEmembership{Fellow,~IEEE}
\thanks{This work has been submitted to the special issue of the IEEE Transactions on Information Theory titled: ``Facets of Coding Theory: from Algorithms to
Networks". This work was supported in part by the NSF Theoretical
Foundation Grant 0728676. Hakim Alhussien was with the department
of Electrical and Computer Engineering, University of Minnesota,
Minneapolis, Minnesota 55455, U.S.A. He is now with Link-A-Media
Devices, Santa Clara, CA 95051, USA
(e-mail:hakima@link-a-media.com). Jaekyun Moon was with the
department of Electrical and Computer Engineering, University of
Minnesota, Minneapolis, Minnesota 55455, U.S.A. He is now with the
department of Electrical Engineering at KAIST, Yuseong-gu,
Daejeon, 305-701, Republic of Korea
(e-mail:jaemoon@ee.kaist.ac.kr).}}

\maketitle

\begin{abstract}
The error-pattern correcting code (EPCC) is incorporated in the
design of a turbo equalizer (TE) with aim to correct dominant
error events of the inter-symbol interference (ISI) channel at the
output of its matching Viterbi detector. By targeting the low
Hamming-weight interleaved errors of the outer convolutional code,
which are responsible for low Euclidean-weight errors in the
Viterbi trellis, the turbo equalizer with an error-pattern correcting code
(TE-EPCC) exhibits a much lower bit-error rate (BER) floor
compared to the conventional non-precoded TE, especially for high rate
applications. A maximum-likelihood upper bound is developed on the
BER floor of the TE-EPCC for a generalized two-tap ISI channel, in order to
study TE-EPCC's signal-to-noise ratio (SNR) gain for various
channel conditions and design parameters. In addition,
the SNR gain of the TE-EPCC relative to an existing precoded TE
is compared to demonstrate the present TE's superiority
for short interleaver lengths and high coding rates.
\end{abstract}



\newpage

\section{Introduction}\label{Intro}

The turbo code of~\cite{Berrou93,Benedetto96} has been utilized as
a practical means to approach the inter-symbol-interference (ISI)
channel capacity in what has been termed turbo
equalization~\cite{Douillard95,Koetter04}, in which two recursive
systematic convolutional codes (RSCCs) concatenated in parallel
are concatenated serially to the ISI channel. Since then, the
turbo equalization terminology has grown to encompass any
soft-decodable code that is iteratively decoded by exchanging soft
information with a channel matched detector. The family of turbo
equalizers now includes low-density-parity check (LDPC) codes and
turbo product codes (TPC). A standard turbo code is a parallel
concatenation of convolutional codes (PCCCs) connected by an
interleaver, for which the probability of generating low Euclidean
weight error events is considerably reduced by the action of the
uniform interleaver. This in effect improves the overall system
bit-error-rate (BER) in the low-to-medium signal-to-noise ratio
(SNR) region. A PCCC is decoded by an iterative exchange of soft
information between maximum \textit{a posteriori} probability
(MAP) decoders matched to the constituent RSCC
decoders~\cite{Hagenauer96}. A turbo equalizer (TE) based on an
iterative receiver composed of a PCCC soft decoder and a channel
detector was discussed in~\cite{Ryan98}. A simpler serial
concatenation of a single RSCC and a precoder through an
interleaver was found to perform just as well in~\cite{Reed98} for
wireless communication applications, and in~\cite{Souvignier00}
and~\cite{McPheters99} for magnetic recording applications.
Precoding makes the ISI channel appear recursive to the outer
interleaved RSCC, where the non-precoded ISI channel can viewed as
an inner nonrecursive rate-$1$ convolutional
code~\cite{Koetter04}. In this manner, precoding is essential to
achieve better turbo gain in the low SNR region, i.e. ``waterfall
region". This was first shown in the context of serially
concatenated convolutional codes (SCCCs) in additive white
Gaussian noise (AWGN) in~\cite{Benedetto98}, where it was
demonstrated that the inner constituent convolutional code has to
be recursive to achieve a turbo gain. Briefly afterwards, this was
demonstrated for a SCCC-TE running on the dicode channel
in~\cite{Oberg01}. The concatenation of precoding and RSCC through
an interleaver works by enhancing the error weight ``spectral
thinning" effect, by which the frequency of low Euclidean distance
errors is uniformly reduced.

We propose an alternate error-weight spectral shaping approach
that aggressively targets the low end of the error Euclidean
distance distribution, enhancing BER performance in the ``error
floor" region, while maintaining the waterfall region gain of
conventional TE. The proposed method is based on directly
targeting the dominant error patterns of the channel, which are
also the lowest Euclidean distance errors, via a matching error
correction code, termed the error-pattern correction code (EPCC).
The EPCC was first proposed to handle single dominant error event
occurrence in~\cite{MoonICC05}~\cite{intermag06}, and later
enhanced in~\cite{intermag07} and~\cite{Jih_Moon} to handle
multiple error event occurrences. A practical EPCC-based turbo
equalizer tailored to the magnetic recording application was first
proposed in~\cite{Hakim07}. In our TE setup, the EPCC is matched
to the non-precoded ISI channel and serves as an inner code for an
outer interleaved RSCC. Since the EPCC maintains a substantial
error correction power while having a high code rate that is close
to $1$, the hope is that the redistribution of redundancy between
EPCC and the outer RSCC would improve overall system performance.
In TE-EPCC decoding, the EPCC MAP decoder works iteratively with
the outer RSCC MAP decoder to correct low Euclidean distance
errors at the output of the channel's detector. This is compared
to using a rate-$1$ precoder in the encoder side that prevents
these errors from occurring in high frequency but can not
eliminate them entirely.

In this work, we conduct an error-event weight analysis of EPCC
enhanced TE to be able to predict an upper bound on the BER
performance, and hence establish the advantage of incorporating
EPCC in the error floor region. The derived upper bound on BER is
for the maximum likelihood (ML) decoder of the concatenated
system, which the practical decoder is assumed to approach at high
SNRs. A few points are worth mentioning regarding the derivation
of such a bound. First, the bound is based on the notion of a
uniform interleaver, which essentially averages out the effect of
good and bad instantaneous interleavers on the bound. The
implication of this assumption on the analytic BER bound is that
the particular choice of the practical interleaver is not a factor
in our turbo system comparison herein. Second, the derivation of
the bound presumes a maximum-likelihood decoder, which fails short
of accurately describing the iterative turbo gain that is more
pronounced at lower to medium SNR, where the analysis of turbo
code performance at this lower SNR region remains largely an open
problem. Incidentally, our proposed approach here based on
probabilistic correction of low Euclidean distance errors is
designed to work in the floor region where the bound is accurate.
Finally, the bound assumes that coded data is i.i.d., which
becomes a more realistic approximation as the code rate of the
RSCC approaches unity.

The paper is organized as follows; In Section \textrm{II} we
review the main concepts of EPCC code construction and decoding
based on its algebraic properties; we also present EPCC design
examples that we later use in the simulation of Section
\textrm{VI}. In Section \textrm{III} we present the encoder and
decoder components of the conventional precoded and non precoded
TEs and of the TE-EPCC. Section \textrm{IV} analyzes the ML BER
performance of the TE-EPCC and the conventional TE based on the
overall error weight spectrum of the coded channel. Furthermore,
this section discusses an efficient method to evaluate the BER
bound based on multinomial theory, assuming a single EPCC codeword
per interleave. In Section \textrm{V} we explain the gain of the
TE-EPCC over the TE in terms of the improved interleaver gain
exponents of lower Euclidean-weight errors. Section \textrm{VI}
discusses a practical method to implement TE-EPCC decoding that
approaches the ideal ML decoder analyzed in the preceding section.
Finally, The numerical results in Section \textrm{VII} corroborate
our claims in a variety of channel conditions for a combination of
decoder design parameters.

\section{Review of the Error-Pattern-Correcting Code}\label{secII}
The cyclic codes described in~\cite{intermag06} are based on
construction of a generator polynomial $g(x)$ that gives rise to
distinct syndrome sets for all targeted dominant error patterns.
It has been shown that such a  $g(x)$ can be obtained from the
irreducible factors making up the polynomial representations of
the dominant error patterns. The code can be further improved by
introducing another factor in $g(x)$, namely, a primitive
polynomial that is not already a factor of
$g(x)$~\cite{intermag07}. The results are an increased code rate,
improved single-error-pattern correction accuracy (via reduced
miss-correction probability), and capability to correct some
important multiple-pattern events based on a increased number of
distinct syndrome patterns.

We start by constructing a cyclic code targeting the set of
$l_{max}$ dominant error events
$$\{e^{(1)}_k(x),e^{(2)}_k(x),...,e^{(l_{max})}_k(x)\}$$
represented as polynomials on $GF(2)$ that can occur at any
starting position $k$ in the codeword of length $l_T$. A syndrome
of error $e^{(i)}(x)$ at position $k$ is defined as $s^{(i)}_k(x)=
e^{(i)}_k(x)\,\mathbf{mod}\,g(x)$ , with $g(x)$ being the
generator polynomial of the code and $\mathbf{mod}$ the polynomial
modulus operation. A syndrome set $\mathbf{S}_i$ for error type
$e^{(i)}(x)$ contains elements corresponding to all cyclic shifts
of polynomial $e^{(i)}(x)$; elements of $\mathbf{S}_i$ are thus
related by $s_{k+j}^{(i)} \equiv x^{j}s_k^{(i)}\mathbf{mod}g(x)$.

For unambiguous decoding of $e^{(i)}(x)$ and $e^{(j)}(x)$,
$\forall\{i,j\}$, we must have $\mathbf{S}_i \cap \mathbf{S}_j =
\oslash$. This design requirement constrains $g(x)$ to have
distinct greatest common divisors with all $e^{(i)}(x)$. However,
even if this constraint is satisfied, an element in $\mathbf{S}_i$
can still map to more than one position, i.e., the period of the
syndrome set- and period of $g(x)$- can be less than $l_{max}$.
Moreover, this constraint is only sufficient but not necessary.
Also, as shown in~\cite{intermag06}, there may exist a lower
degree $g(x)$ that can yield distinct syndrome sets for the
targeted error polynomials, resulting in a higher rate EPCC. A
search method to find this $g(x)$ is already discussed in detail
in ~\cite{intermag06} and~\cite{Jih_Moon}.

We now describe the construction and properties of the EPCC that
will be deployed throughout the paper in the design of different
turbo systems based on EPCCs. We target the dominant error events
of a generalized two tap ISI channel of the form $1-\alpha D,\
0<\alpha\leq 1$, for which the dicode and PR$1$ channels are
special cases. When $\alpha$ is close to $1$, the dominant errors
are: $+$, $+-$, $+-+$, etc., which have the polynomial
representations: $e^{(1)}(x)=1$, $e^{(2)}(x)=1+x$,
$e^{(3)}(x)=1+x+x^2$, etc., i.e. polynomials on $GF(2)$ for which
all powers of $x$ have nonzero coefficients.

For the purpose of designing EPCC codes for use in the TE-EPCC,
the component EPCC code rate should be very high. To maintain high
rate, the EPCC codeword has to be extended to a few hundred bits,
without proportionally increasing the number of parity bits
required to achieve accurate single-error occurrence correction
capability. Example EPCC codes are shown next, and the syndrome
set periods of these codes are shown in
Table~\ref{Table:SyndromePeriod}.

\begin{itemize}
    \item \emph{$(630,616)$ EPCC}: Targeting error polynomials up to degree $9$, we get the generator polynomial $g(x)=1 +x^3 +x^5 +x^8$ of period
    $30$, via the search procedure in~\cite{intermag06}. Choosing a codeword length of $30$, $10$ distinct, non-overlapping syndrome sets are utilized to
    distinguish the $10$ target errors. However, the resulting $(30,22)$ EPCC has rate $0.73$ which incurs high rate
    penalty. By multiplying the base EPCC generator polynomial by the
    primitive polynomial $1+x+x^6$, which is not a factor of any of the targeted errors, we obtain the extended
    generator polynomial $g_e(x)=1+ x+ x^3+ x^4+ x^5+ x^8+ x^{11}+
    x^{14}$, which corresponds to the extended $(630,616)$ EPCC code
    of rate $0.98$, and $14$ parity bits. Then, as shown in~\cite{intermag06}, syndrome
    sets $\mathbf{S}_1$, $\mathbf{S}_3$, $\mathbf{S}_7$, and $\mathbf{S}_9$ have period $630$ and thus can
    be decoded without ambiguity. On the other hand, syndrome sets $\mathbf{S}_2$, $\mathbf{S}_4$, $\mathbf{S}_6$, and $\mathbf{S}_8$ have period
    $315$, decoding to one of two positions. The worst would be
    $\mathbf{S}_5$ of period $126$, and $\mathbf{S}_{10}$ of period
    $63$, which decode to $5$ and $10$ possible positions, respectively. Still, the algebraic decoder can quickly shrink the number of possible error positions to few
    positions by checking the data support, and then would choose the
    one position with highest local reliability.
    \item \emph{Shortened $(126,112)$ EPCC}: Shorter lower-rate EPCC codes can be obtained by shortening the $(630,616)$ EPCC. For example, a
    $(126,112)$ EPCC of rate $0.89$ can be derived this way with all syndromes sets, excluding syndrome set
    $\mathbf{S}_{10}$, having period $126$, and thus are decodable without ambiguity.
    \item \emph{$(210,199)$ EPCC}: To obtain short EPCC codes without jeopardizing the code rate through code shortening, we can target fewer error patterns in the
    code design. Targeting error polynomials up to
    degree $9$, but excluding $e^{(7)}(x)$, we can extend the base generator polynomial $g(x)=1 +x^3 +x^5
    +x^8$ through its multiplication by the primitive polynomial $1+x+x^3$,
    which we could not use before because its a factor of the
    polynomial representation of $e^{(7)}(x)$. The resulting code
    is a $(210,199)$ EPCC of rate $0.95$, $11$ parity bits, and extended generator
    polynomial $g_e(x)=1+x+x^4+x^5+x^9+x^{11}$.
\end{itemize}

\begin{center}
\begin{table} [!ht] 
    \caption{Syndrome set periods of various EPCC codes.}
    \label{Table:SyndromePeriod}
    \centering
    \begin{tabular}{|l||l|l|l|} \hline
         \begin{sideways} Target error \end{sideways}          & \begin{sideways} $(630,616)$ EPCC \ \ \end{sideways} & \begin{sideways} $(126,112)$ EPCC \ \ \end{sideways} &  \begin{sideways} $(210,199)$ EPCC \ \ \end{sideways}   \\  \hline\hline
        $1$                   & $630$          & $126$           & $210$\\  \hline
        $(1 + x)$             & $315$          & $126$           & $105$\\  \hline
        $(1 + x + x^2)$       & $630$          & $126$           & $70$ \\  \hline
        $(1 + x)^3 $          & $315$          & $126$           & $105$\\  \hline
        $(1 + x + x^2 + x^3 + x^4)$            & $126$           & $126$     & $42$  \\  \hline
        $(1 + x)(1 + x + x^2)^2$               & $315$           & $126$     & $35$  \\  \hline
        $(1 + x + x^3)(1 + x^2 + x^3)$         & $630$           & $126$     & $-$   \\  \hline
        $(1 +x)^7$                             & $315$           & $126$     & $105$ \\  \hline
        $(1 + x + x^2)(1 + x^3 + x^6)$         & $630$           & $126$     & $70$  \\  \hline
        $(1 + x)(1 + x + x^2 +x^3 + x^4)^2$    & $63$            & $63$      & $21$  \\  \hline
    \end{tabular}\\
\end{table}
\end{center}

\section{A TE Incorporating the EPCC SISO Decoder}\label{secIII}
\begin{figure}[hbtp]
\centering
\includegraphics*[width=\figsz in]{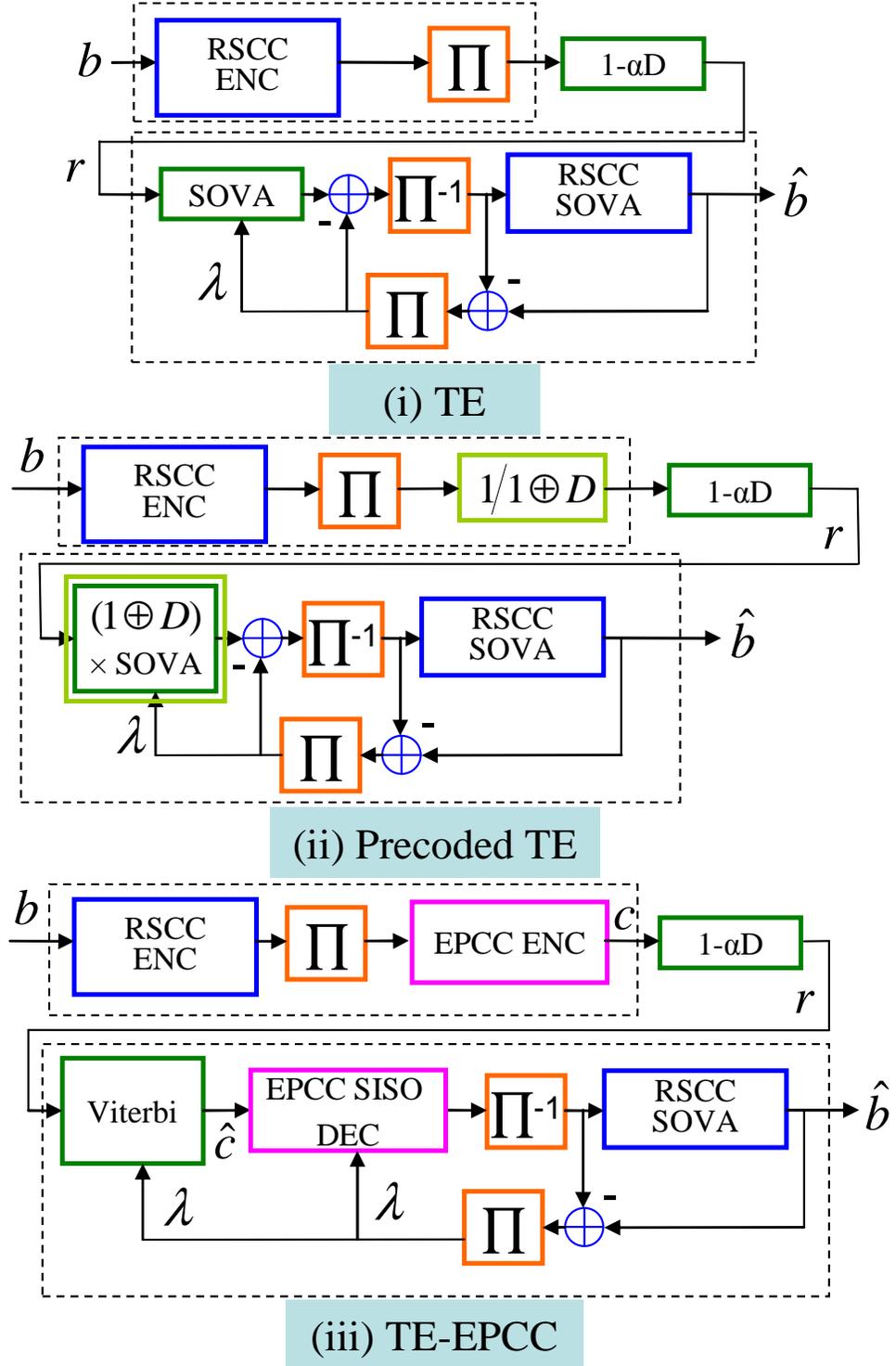}
\caption{Block diagrams: (\textrm{i}) TE, (\textrm{ii}) precoded
TE, (\textrm{iii}) TE-EPCC.}\label{TEEPCC_blk}
\end{figure}

The structure of the conventional SCCC-TE is shown in
Fig.~\ref{TEEPCC_blk}(\textrm{i}). In the encoder side, a simple
RSCC encodes the data stream, which is interleaved before being
passed to the channel. The concatenation of the convolutional code
and ISI channel can be viewed in the context of turbo coding as a
serial concatenation of an outer recursive code and an inner
rate-$1$ non-recursive code through an interleaver. A
polynomial-time iterative-type decoder can be designed based on
the separation of the ML decoders of the inner and outer codes.
The ML decoders iteratively exchange reliability information
converging to the combined ML solution at high SNR. Separate ML
detection and decoding can be realized via the
Bahl-Cocke-Jelinek-Raviv (BCJR) algorithm~\cite{BCJR}, the soft
output Viterbi algorithm (SOVA)~\cite{Hagenauer89}, or the minimum
mean squared error (MMSE) soft-in soft-out (SISO)
detector~\cite{Tuchler02com,Tuchler02sp}. The BER gain in TE,
however, is most notable at low SNRs, and declines rapidly as SNR
increases resulting eventually in the error floor phenomenon. The
gain at low SNR is further enhanced by including a rate-$1$
recursive component in the path of the coded interleaved bit
stream. This is shown in Fig.~\ref{TEEPCC_blk}(\textrm{ii}), where
the trellis of SOVA is now matched to the recursive rate-$1$ coded
channel $\frac{1 - \alpha D}{ 1 \oplus D}$. By the action of the
ideal uniform interleaver, the fraction of errors in the
Hamming-weight error distribution of the RSCC resulting in low
Euclidean weight errors in the channel trellis is greatly reduced.
This, as a result, improves the BER at low to medium SNR, where
the contribution of the profile of error Euclidean weights to the
BER far exceeds the single contribution of the minimum of these
weights.

A markedly different approach is proposed in the structure of
Fig.~\ref{TEEPCC_blk}(\textrm{iii}). The new method is based
on replacing the rate-$1$ precoder with a high rate ECC that is
designed to correct low Hamming weight errors that generate low
Euclidean-weight trellis errors rather than constraining their
incidence. Since the targeted errors possess low Hamming weights
by design, this reduces the added complexity of encoding and
decoding the EPCC, while the intrinsic channel property of these
errors generating low Euclidean weight errors, particulary
$d_E^2=2$, lowers the error floor at medium to high SNRs
substantially. Nevertheless, since the practical decoder of the EPCC
incurs some miscorrection, this new approach resembles a
probabilistic ``best effort" enhancement of $d_{E, min}^2$ that is
achieved by correcting a sizable fraction of the originating
Hamming weight errors. A soft-in soft-out (SISO) decoder of the EPCC
is assumed in the iterative turbo loop. Since the EPCC is matched to
the ISI channel, no interleaving should be present between the EPCC
and the channel. On the other hand, an interleaver is essential
between the EPCC and the outer RSCC.

While the $\frac{1}{1 \oplus D}$ precoder and the $1- \alpha D$
channel can be jointly decoded with no added complexity by
matching the trellis to the combined coded channel $\frac{1 -
\alpha D}{1 + \oplus D}$, its impractical to realize a similar
joint ML decoder of the channel and the EPCC. Hence, in
Fig.~\ref{TEEPCC_blk}(\textrm{iii}) separate SISO decoders of
the channel and the  EPCC are implemented.

\section{Error-Rate Analysis of TE-EPCC}

In bounding the BER of the TE-EPCC, many of the basic steps
and assumptions taken in
\cite{Oberg01} and \cite{Oberg01ICC} for bounding the BER of the conventional TE
are utilized. To more closely reflect the practical recording channel,
however, we apply our BER analysis to
a generalized two-tap channel of the form $1 \pm \alpha D$. The
dicode ($1-D$) and PR1 ($1+D$) channels are special cases corresponding to
$\alpha=1$. In the proposed approach, we show how the BER is
function of the error Euclidean distance distribution of the
overall system. Then, we argue for TE-EPCC's enhanced performance by
the virtue of its reduction of occurrence frequencies of low
Euclidean distances in the overall distance distribution; it
will also be shown that these low Euclidean distance components of
the distribution dominate the system BER. Following the notations
of \cite{Oberg01}, the maximum likelihood (ML) union bound on word
error rate of a block code of codebook size $M$, of equally likely
codewords and AWGN of zero mean and variance $\sigma^2$
is
\begin{equation}\label{eqIV1}
    P_W \leq \frac{1}{M}
    \sum_{m=1}^{M}\sum_{\acute{m}\neq m}Q\left(\frac{\parallel\mathbf{x}_m-\mathbf{x}_{\acute{m}}\parallel}{\sigma}\right)
\end{equation}
where $m$ and $\acute{m}$ are codewords separated by the Euclidean
distance $\parallel\mathbf{x}_m-\mathbf{x}_{\acute{m}}\parallel$,
and $\mathbf{x}_m$ is the noiseless channel output for $m$. If
there are $T_{m,d_E}$ different codewords for which the corresponding noiseless channel
outputs are at distance $d_E$ from
$\mathbf{x}_{m}$, then we can write (\ref{eqIV1}) as:
\begin{eqnarray}\label{eqIV2}
  \nonumber P_W &\leq& \frac{1}{M} \sum^{M}_{m=1}\sum_{d_E=1}^{\infty} T_{m,d_E} Q\left(\frac{d_E}{\sigma}\right) \\
   &=& \sum_{d_E=d^{min}_E}^{\infty} \overline{T}(d_E) Q\left(\frac{d_E}{\sigma}\right)
\end{eqnarray}
where $\overline{T}(d_E)$ is the average number of codewords at Euclidean distance
$d_E$ from a given codeword, with the distance measured at the channel output.
The
associated BER can be shown to be
\begin{equation}\label{eqIV3}
  P_b \leq \sum_{d_E=d^{min}_E}^{\infty} \frac{\overline{T}(d_E)\overline{w}(d_E)}{K} Q\left(\frac{d_E}{\sigma}\right)
\end{equation}
where $K$ is the number of information bits per codeword sequence
and $\overline{w}(d_E)$ is the average Hamming distance from a given information
word to competing information words located at $d_E$ away, with the Euclidean distance
measure based on noiseless channel outputs of the corresponding codewords.
We next show how $\overline{T}(d_E)$ is
related to the outer code Hamming weight enumerator
$\mathbf{A}^o(d)$ and the error event characteristics of the
channel.

\subsection{Error Event Analysis of the $1 - \alpha D$ Channel} 
\begin{figure}[hbtp]
\centering
\includegraphics*[width= 4 in]{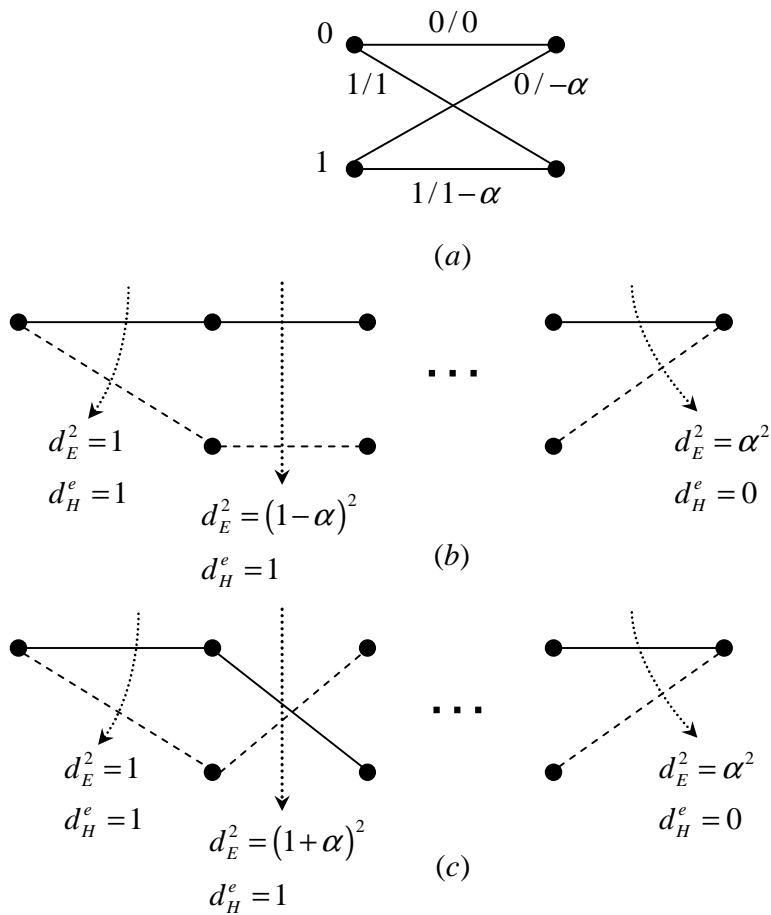}
 \caption{(a) Trellis section for a non-precoded generalized two-tap ISI
channel($1-\alpha D$), (b) dominant error patterns, (c)
non-dominant error patterns.}\label{nonprecoded_Tr}
\end{figure}

A
trellis section of the $1 - \alpha D$ channel with no precoding is
shown in Fig.~\ref{nonprecoded_Tr}. The branch label $c_i/x_i$ signifies the
coded input bit to the channel, and the corresponding channel
output, respectively. Following the same notation
as in \cite{Oberg01}, any error word $\mathbf{f}$ with Hamming
weight $d=d^H(\mathbf{f})$ can be uniquely decomposed into a
concatenation of disjoint error patterns $\mathbf{f}_j$,
$j=1,\ldots,m$, where the index $j$ signifies the order of
occurrence of the error pattern of Hamming weight $d^H_j$ in the
codeword. Error patterns $\mathbf{f}_j$, $j < m$, correspond to
simple closed error events on the trellis that diverge from and
remerge into the correct path without sharing any of the states in
between. However, two scenarios can occur when $j=m$: either
$\mathbf{f}_m$ remerges with the correct path (closed
$\mathbf{f}_m$) or the boundary of the codeword is reached while
the two paths are still diverged (open $\mathbf{f}_m$).

In the $1 - \alpha D$ channel trellis, diverging branches result
in a Euclidean distance separation of $1$ each, while remerging
branches result in a squared Euclidean distance separation of
$\alpha^2$ each. Moreover, crossing branches accumulate a squared
distance separation of $(1+\alpha)^2$, while parallel branches
accumulate a separation of $(1-\alpha)^2$. This means that
parallel branches result in a lower Euclidean distance separation
compared to crossing branches in the Euclidean distance
distribution when $0 < \alpha \leq 1$.

Hence, two error pattern classes are distinguishable according to
their accumulate Euclidean distance. The first class, shown in
Fig.~\ref{nonprecoded_Tr}b, has a squared distance
$d_E^2(\mathbf{f}_j)=1+\mu\alpha^2+(d^H_j-1)\times(1-\alpha)^2$
where $d^H_j$ is the Hamming weight of the error event
$\mathbf{f}_j$, and $\mu$=0 or 1 depending on the event being open or closed,
respectively. This class of error patterns is denoted by
$\chi^{dom}$ and is called ``the dominant error class", for which
all branches, other than the diverging and remerging branches, are
parallel. The dominant error class accounts for most of the
channel bit errors due to the low Euclidean distance between the
correct and erroneous paths. On the other hand, the second class,
shown in Fig.~\ref{nonprecoded_Tr}c, has both parallel and
crossing branching, and hence its members have Euclidian distance
$d_E^2(\mathbf{f}_j)=1+\mu\alpha^2+\lambda_{cr} \times
(1+\alpha)^2+(d^H_j-1-\lambda_{cr})\times(1-\alpha)^2$, where
$\lambda_{cr}$ is the number of crossing branches. The second
class contributes much less to the overall system BER, and thus we
call it ``the non-dominant error class", which is denoted by
$\widetilde{\chi}^{dom}$. By the same line of argument, the same two
classes are distinguishable for the PR1 channel, which is a
special case of $1 + \alpha D$ at $\alpha=1$. The only difference
is that error events with all crossing branches now generate the
class $\chi^{dom}$.

\begin{figure}[hbtp]
\centering
\includegraphics*[width= \figsz in]{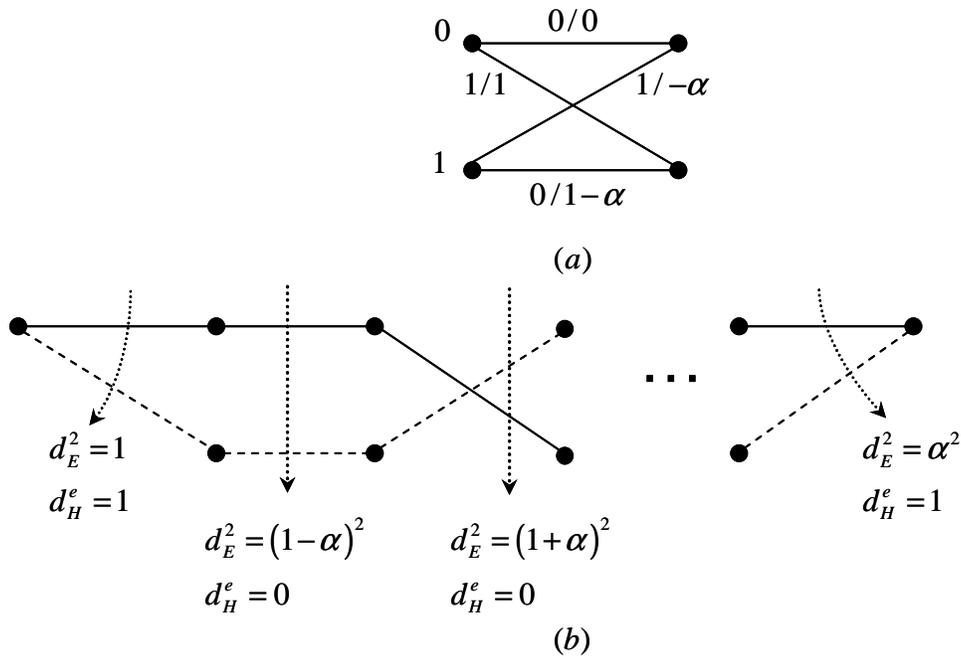}
 \caption{(a) Trellis section for a $\frac{1}{ 1 \oplus D}$ precoded generalized two-tap ISI
channel ($1-\alpha D$), (b) weight characterization of an error
pattern}\label{precoded_Tr}
\end{figure}
We design an error-pattern correcting code (EPCC) capable of
correcting error codewords $\mathbf{f}$ that are decomposable into
disjoint error patterns $\mathbf{f}_j$ that all belong to the
dominant error class, i.e. $\mathbf{f}_j\in\chi^{dom},\forall j$.
In order to evaluate the BER performance of EPCC we need to find
the new Euclidean distance distribution modified by EPCC. However,
it would be easier to first find the Euclidean distance
distribution before EPCC correction is turned on. We assume
throughout that code bit values are i.i.d and equiprobable, which
is a valid assumption for high rate codes. Suppose an error word
$\mathbf{f}$, of Hamming weight $d^H(\mathbf{f})=d$, is composed
of $m_{dom}$ error patterns $\mathbf{f}_j\in\chi^{dom}$, and
$\widetilde{m}_{dom}=m-m_{dom}$ error patterns $\mathbf{f}_j\in
\widetilde{\chi}^{dom}$. A dominant error pattern $\mathbf{f}_j$
of length $l_j=d_H(\mathbf{f}_j)$ will have probability
$\left(\frac{1}{2}\right)^{l_j-1}$. On the other hand, a
non-dominant error pattern $\mathbf{f}_j$ of length $l_j$ and
$\lambda_{cr}$ crossing branches will have a probability of $
{{l_j-1} \choose {\lambda_{cr}}}
\left({\frac{1}{2}}\right)^{l_j-1}$. Therefore, the probability
distribution of $d^2_E(\mathbf{f})$ is given by:
\begin{eqnarray}\label{eqIVA1}
\nonumber \textrm{Pr}(d_E| d,m) & = & \left \{ \begin{array}{ll}
 { {d-m}  \choose {\lambda_{cr}} }\left( {\frac{1}{2}}\right)^{d-m}, & \begin{array}{l} \lambda_{cr}>0 \ \textmd{integer}, \\ \ m_{dom} < m. \end{array} \\
 \left({\frac{1}{2}}\right)^{d-m_{dom}}, & \begin{array}{l} \lambda_{cr}=0, \\ \ m_{dom} = m. \end{array} \\
 0, & \textmd{otherwise.}
\end{array}
\right. \\
\lambda_{cr} & = & \frac{d_E^2 - (1-\alpha)^2 d - 2 \alpha m + \mu
\alpha^2}{4 \alpha}
\end{eqnarray}
which is the conditional probability of an error word of Euclidean
distance $d^2_E$, given that its Hamming weight is $d$, and has
$m$ multiple error pattern occurrences, of which $m_{dom}$ belong
to $\chi^{dom}$.

If we examine the precoded $1-\alpha D$ trellis in
Fig.~\ref{precoded_Tr}, we note that a nonzero Hamming error
results in the diverging of a single error event that remerges
only on the occurrence of another Hamming error, while all the in
between error branches have zero Hamming weights, wether crossing
or parallel. We also note that an even $d^H$ compound Hamming
error decomposes into $\frac{d^H}{2}$ closed single errors, while
an odd $d^H$ compound error decomposes into $\lfloor \frac{d^H}{2}
\rfloor$ closed errors and a boundary error. Moreover, diverging
and remerging branches have $d_E^2=1$ and $d_E^2=\alpha^2$,
respectively, while parallel and crossing branches have
$d_E^2=(1-\alpha)^2$ and $d_E^2=(1+\alpha)^2$, respectively. This
means that, by invoking the random uniform interleaver assumption,
the probability of a single long error event of $d^H=\{1, 2\}$
producing a low Euclidean weight error declines rapidly as the
interleaver length is increased, since the probability of an all
parallel error event declines accordingly. The Euclidean distance
of a multiple error event of Hamming weight $d$, $\lambda_{cr}$
crossing branches, and total length $L$ is:
\begin{eqnarray}\label{eqIVA2}
\nonumber d_E^2= \lceil \frac{d}{2} \rceil +
\lfloor\frac{d}{2}\rfloor \alpha^2 + (1+ \alpha)^2 \lambda_{cr} +
(1-\alpha)^2(L-d- \lambda_{cr})
\end{eqnarray}
Therefor
\begin{eqnarray}\label{eqIVA3}
\nonumber \textrm{Pr}(d_E| d,L) & = & \left \{ \begin{array}{ll}
 { {L-d}  \choose {\lambda_{cr}} }\left( {\frac{1}{2}}\right)^{L-d}, & \lambda_{cr}>0 \ \textmd{integer}. \\
 0, & \textmd{otherwise.}
\end{array}
\right. \\
\nonumber \lambda_{cr} & = & \frac{d_E^2 - \lceil \frac{d}{2}
\rceil - \lfloor\frac{d}{2}\rfloor \alpha^2 - (1-\alpha)^2(L-d)}{4
\alpha}
\\
\end{eqnarray}
\begin{figure}[hbtp]
\centering
\includegraphics*[width= \figsz in]{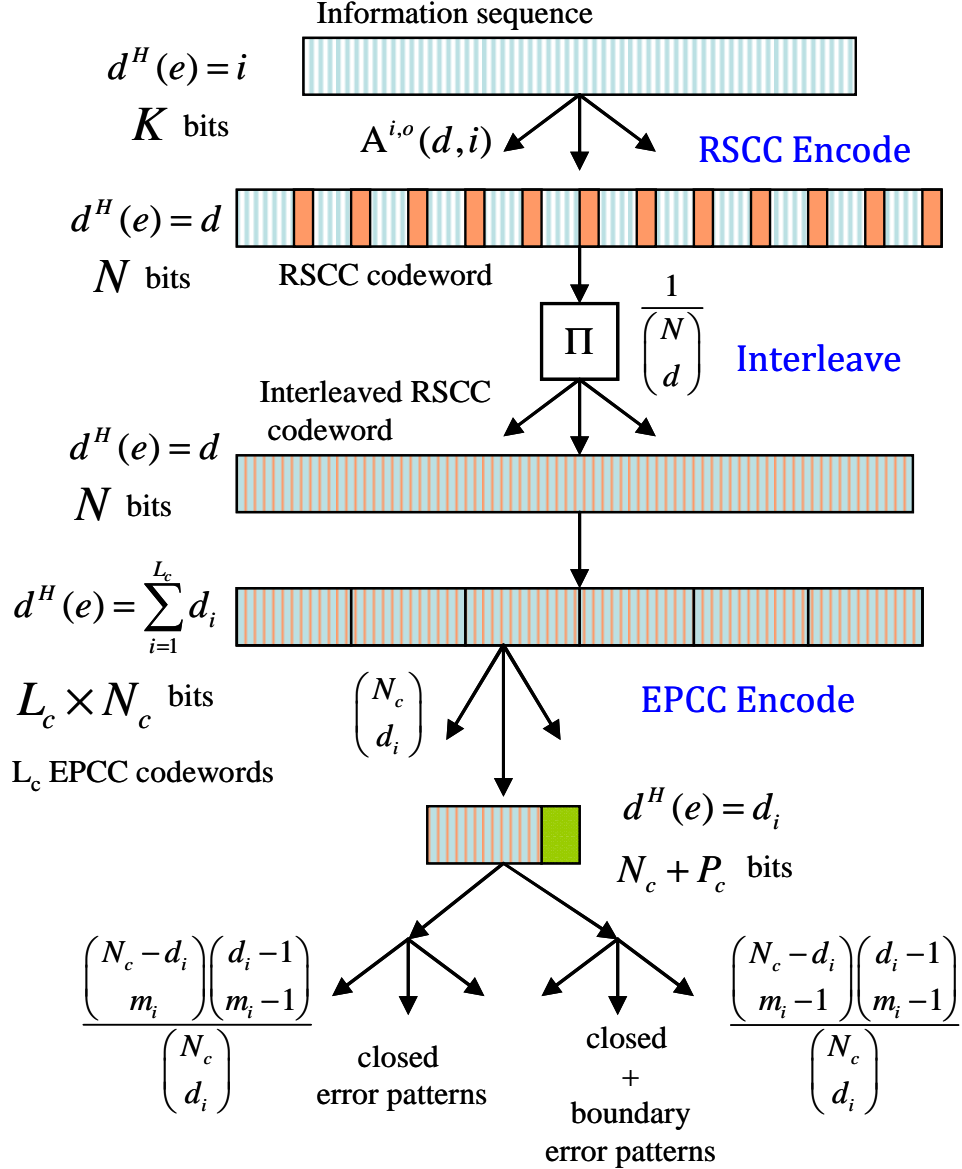}
\caption{Sketch of the method to derive
$\overline{T}(d_E)$.}\label{proof_sketch}
\end{figure}
\subsection{Error Euclidean Distance Distribution of TE-EPCC}
 We now develop a method to construct the error Euclidean distance
distribution of TE-EPCC, for which the comparable distance
distribution of TE is a special case where EPCC is turned off.
Consider a serial concatenation of an EPCC and an interleaved
recursive systematic convolutional code (RSCC) of length $N$.
There are $L_c$ EPCC subcodes in each interleave, each of length
$N_c=\frac{N}{L_c}$, where an EPCC can correct up to $m_c$
multiple occurrences per subcode provided that they all belong to
the target set of correctable errors. The target set is
$\{\mathbf{f}_j:\ \mathbf{f}_j\in\chi^{dom},\ d^H(\mathbf{f}_j)
\leq d_c \}$, where $d_c$ is the maximum length correctable error
from $\chi^{dom}$. An error in the RSCC codeword of hamming weight
$d$ is mapped by the uniform interleaver into all possible ${ N
\choose d }$ interleaved error words $\mathbf{f}$ with equal
probability. The interleaved error word divides into $L_c$ EPCC
subcodes, each receiving error word $\mathbf{f}^{(i)}$, $i= 1,
...,L_c$, of Hamming weights $d_1,\ d_2,\ ...,\ d_{L_c}$. Each
EPCC error word $\mathbf{f}^{(i)}$ of Hamming weight $d_i$
decomposes into $m_i$ disjoint error pattern occurrences. In the
previous section, we found the conditional probability $P(d_E|
d,m)$ given the error Hamming weight and number of multiple errors
$m$ for a single subcode interleave. To derive the Euclidean
distance distribution for a codeword that is divisible into $L_c$
subcodes, we are also required to evaluate the conditional
probability of the decompositions $m_i$ given the EPCC subcode
hamming weights $d_i$. The conditional Euclidean distance
probability distribution can be expanded as follows:
\begin{eqnarray}\label{eqIVB1}
  \nonumber \textrm{Pr}(d_E| d) &=& \textrm{Pr}(d_E| d,d_1, ..., d_{L_c})\times \textrm{Pr}(d_1, ..., d_{L_c}| d) \\
  \nonumber  &=& \textrm{Pr}(d_E| d,d_1, ..., d_{L_c}, m, m_1, ..., m_{L_c})\\
  \nonumber & & \times \textrm{Pr}(m,m_1, ..., m_{L_c}| d_1, ..., d_{L_c},d) \\
            & & \times \textrm{Pr}(d_1, ...,
   d_{L_c}| d).
\end{eqnarray}
Since errors in the $L_c$ EPCC subcodes are disjoint,
(\ref{eqIVB1}) becomes:
\begin{eqnarray}\label{eqIVB2}
\nonumber \textrm{Pr}(d_E| d) & = & \mathop
{\sum_{d_1=0}^{d}...\sum_{d_{L_c}=0}^{d}}\limits_{d=\sum_{i=1}^{L_c}
d_i} \textrm{Pr}(d_1,...,d_{L_c}| d) \\
\nonumber & & \sum_{m=1}^{d}\mathop
{\sum_{m_1=0}^{d_1}...\sum_{m_{L_c}=0}^{d_{L_c}}}\limits_{m=\sum_{i=1}^{L_c}
m_i} \textrm{Pr}(d_E| d,m)\prod_{i=1}^{L_c}\textrm{Pr}(m_i| d_i)
\\
\end{eqnarray}
The joint conditional probability $\textrm{Pr}(d_1,...,d_{L_c}|
d)$ in (\ref{eqIVB2}) is the probability of dividing the ${ N
\choose d }$ possible instants of the interleaved error word
$\mathbf{f}$, of Hamming weight $d$, into the error word sequence
$\mathbf{f}^{(i)}$ with associated Hamming weight sequence $d_i$,
and is given by
\begin{equation}\label{eqIVB3}
\textrm{Pr}(d_1,...,d_{L_c}| d)=\frac{{ N_c \choose d_1 } \times {
N_c \choose d_2 } ...\times { N_c \choose d_{L_c} }}{{ N \choose d
}}.
\end{equation}
Given that there are $d_i$ errors in EPCC subcode $i$, there
exists ${ d_i-1 \choose m_i-1 }$ ways by which the Hamming weight
$d_i$ error is decomposed into $m_i$ multiple error pattern
occurrences, each of length at least $1$. Of these $m_i$
occurrences, $\mathbf{f}^{(i)}_{m_i}$ can be either open or
closed. An open error event in this context lies on the boundary
of the EPCC subcode's data and parity fields. A boundary error
event is defined this way since we do not count error patterns in
the EPCC codeword's parity towards the total BER, where this field
is discarded before passing the decoded data to the outer
interleaved RSCC decoder. By examining the trellis we note that
boundary errors contribute a squared Euclidean distance separation
that is $\alpha^2$ less than identical length closed errors that
are totally encapsulated by the current subcode $i$ data field.
Furthermore, there are only ${ N_c-d_i \choose m_i-1 }$ ways by
which the disjoint $m_i$ error patterns of error word
$\mathbf{f}^{(i)}$ can be arranged in the current subcode $i$,
given the subcode has a boundary error. Two disjoint error
occurrences in the trellis are separated at least by the error
free distance of the channel, which equals $1$ for $1 \mp \alpha
D$ ISI channels. The number of possible arrangements of $m_i$
errors is computed given the fact that the last error pattern
occurs at the boundary. Assuming errors can occur on and off the
boundary, the total number of possible error pattern arrangements
becomes ${ N_c-d_i+1 \choose m_i }$. Given that the EPCC parity
field is long enough, boundary errors have very low probability of
spanning the data fields of adjacent EPCC subcodes, and hence,
such events are independent among different subcodes.

So, given $\mu_i$, there are ${ N_c-d_i \choose m_i-\mu_i }$ ways
by which the $m_i$ error patterns, composing $\mathbf{f}^{(i)}$,
can be arranged in a subcode $i$, and since there are ${ N_c
\choose d_i }$ possible error words $\mathbf{f}^{(i)}$, we get
\begin{equation}\label{eqIVB4}
\textrm{Pr}(m_i| \mu_i, d_i)=\frac{{ N_c-d_i \choose m_i-\mu_i }
\times { d_i-1 \choose m_i-1 } }{{ N_c \choose d_i }}.
\end{equation}
A pictorial depiction of the derivation method explained above is
shown in Figure~\ref{proof_sketch}. Substituting (\ref{eqIVA1}),
(\ref{eqIVB4}), and (\ref{eqIVB3}) into (\ref{eqIVB2}), we get an
expression for the distribution of error Euclidean distances while
EPCC is turned off as:
\begin{eqnarray}\label{eqIVB5}
\nonumber  \textrm{Pr}(d_E| d) &=& \mathop{\sum_{m=1}^{d}
\sum_{\mu=1}^{L_c}}\limits_{
\begin{array}{c} d=\sum_{i=1}^{L_c} d_i,m=\sum_{i=1}^{L_c} m_i,\mu=\sum_{i=1}^{L_c}
\mu_i \\ {m, d, \mu, \alpha}:\ \frac{d_E^2-2 \alpha m+ \mu
\alpha^2 - (1-\alpha)^2 d}{\alpha} =0\ \textrm{mod}\ 4
\end{array}} \\
\nonumber & & \frac{1}{{ N \choose d }}
\prod_{i=1}^{L_c}\sum_{d_i=0}^{d}\sum_{m_i=0}^{d_i}\sum_{\mu_i=0}^{1}
{ d-m \choose \frac{d_E^2-2 \alpha m+ \mu \alpha^2 - (1-\alpha)^2
d}{4\alpha} } \\ \nonumber & & {\left(\frac{1}{2}\right)}^{d-m}{
N_c-d_i \choose
m_i-\mu_i } { d_i-1 \choose m_i-1 } \\
\end{eqnarray}
where we define ${ 0 \choose 0 }=1$. In addition, the Euclidean
distance distribution can be decomposed into two components: a
component $\textrm{Pr}(d_E| d,\mathcal{C})$ associated with error
words that are correctable by the $L_c$ EPCC subcodes, and the
complimentary component $\textrm{Pr}(d_E|
d,\widetilde{\mathcal{C}})$ associated with non-correctable error
words. In this case, the Euclidean distance probability
distribution of non-correctable error words escaping TE-EPCC is
given by
\begin{equation}\label{eqIVB6}
\textrm{Pr}(d_E| d,\widetilde{\mathcal{C}}) = \textrm{Pr}(d_E| d)
- \textrm{Pr}(d_E| d,\mathcal{C})
\end{equation}
while the correctable component is given by:
\begin{eqnarray}\label{eqIVB7}
\nonumber  \textrm{Pr}(d_E| d,\mathcal{C}) &=&
\mathop{\sum_{m=1}^{d} \sum_{\mu=1}^{L_c}}\limits_{
\begin{array}{c} d=\sum_{i=1}^{L_c} d_i,m=\sum_{i=1}^{L_c} m_i,\mu=\sum_{i=1}^{L_c}
\mu_i \\ {m, d, \mu, \alpha}:\ d_E^2=2 \alpha m- \mu \alpha^2 +
(1-\alpha)^2 d \end{array}} \\
\nonumber & & \frac{1}{{ N \choose d }}
\prod_{i=1}^{L_c}\sum_{d_i=0}^{\min(d,d_c)}\sum_{m_i=0}^{\min(d_i,m_c)}\sum_{\mu_i=0}^{1}
\\ \nonumber & & {\left(\frac{1}{2}\right)}^{d-m}{ N_c-d_i \choose m_i-\mu_i }
{ d_i-1 \choose m_i-1 }\\
\end{eqnarray}
where for the sake of simplicity, we assumed that an EPCC subcode $i$ could correct an
error word $\mathbf{f}^{(i)}$ if $d_H(\mathbf{f}^{(i)}) \leq d_c$,
which is actually a worst case scenario that occurs only if
$m_i=1$. Although this assumption would result in a slightly
pessimistic prediction of the EPCC correction power, it allows us to avoid
a substantially more complicated derivation. To obtain the
bound on the bit error probability, we need to express the error
Euclidean distance enumerators as a function of the error
Euclidean distance probability distribution given by
(\ref{eqIVB6}). We note that the average Euclidean weight
enumerator associated with the uncorrectable set of error words
$\widetilde{\mathcal{C}}$ is given by:
\begin{equation}\label{eqIVB8}
\overline{T}(d_E,\widetilde{\mathcal{C}})=\sum_{d=d_{min}}^{N}\mathbf{A}^o(d)\textrm{Pr}(d_E|
d,\widetilde{\mathcal{C}})
\end{equation}
while the average information input hamming distance to codewords
at squared Euclidean distance $d^2_E$ is given by:
\begin{equation}\label{eqIVB9}
\overline{w}(d_E,\widetilde{\mathcal{C}})=\frac{1}{\overline{T}(d_E,\widetilde{\mathcal{C}})}
\sum_{d=d_{min}}^{N}\mathbf{A}^o(d)\mathbf{\overline{A}}^i(d)\textrm{Pr}(d_E|
d,\widetilde{\mathcal{C}})
\end{equation}
where $\mathbf{A}^o(d)$ represents the number of RSCC codeword
sequences of weight $d$, and $\mathbf{\overline{A}}^i(d)$
represents the average input Hamming weight of RSCC codewords of
weight $d$, and are related by
\begin{equation}\label{eqIVB11}
\mathbf{\overline{A}}^i(d)=\frac{\sum_{i}
i\mathbf{A}^{i,o}(d,i)}{\mathbf{A}^o(d)}.
\end{equation}
where $\mathbf{A}^{i,o}(d,i)$ is the number of codeword sequences
of weight $d$ that originated from weight $i$ information
sequences. Details on how to find these marginal error weight
enumerators can be found in~\cite{HakimPhD} for different
puncturing rates and encoder connection polynomials. By
substituting $\overline{T}(d_E,\widetilde{\mathcal{C}})$, given by
(\ref{eqIVB8}), and $\overline{w}(d_E,\widetilde{\mathcal{C}})$,
given by (\ref{eqIVB9}), in (\ref{eqIV3}), we get an upper bound
on the average BER of TE-EPCC as function of $\textrm{Pr}(d_E|
d,\widetilde{\mathcal{C}})$:
\begin{equation}\label{eqIVB10}
 P_b \leq \sum_{d_E=d^{min}_E}^{\infty}\sum_{d=d_{min}}^{N}
 \frac{\mathbf{A}^o(d)\mathbf{\overline{A}}^i(d)\textrm{Pr}(d_E|
d,\widetilde{\mathcal{C}})}{K} Q\left(\frac{d_E}{\sigma}\right).
\end{equation}

In Appendix A we show how these bounds simplify for the simple
case when $L_c=1$, i.e. employing one EPCC subcode per interleave.
Also, we extend the BER bound derived in~\cite{Oberg01} for the
precoded dicode channel to the generalized case $1- \alpha D$.
Finally, by using an exponential-type approximation of the Q
function, we show in appendix A that the BER bounds of the
TE-EPCC, the non-precoded TE, and the precoded TE can be expressed
as single infinite sums, with the Hamming weight of the RSCC error
as the sum index.
\subsection{Efficient Computation of the Euclidean Distance
Enumerator for $L_c>1$ EPCC} A more compact and efficient method
is derived here to evaluate the multiple summations in equations
(\ref{eqIVB7}) and (\ref{eqIVB5}), which are used to compute the
BER bound in (\ref{eqIVB10}). We first define a probability
enumerator for subcode $i$ for all possible values of the
parameters $d_i$, $m_i$ and $\mu_i$, which is given by the
multinomial
\begin{eqnarray}\label{eqIVC1}
\nonumber \mathbf{\Lambda}(\mathrm{D}, \mathrm{M},
\mathrm{\Upsilon};\ m_{max}, d_{max}) = & &
\\ \nonumber 1 +
\sum_{\mu_i=0}^{1}\sum_{d_i=1}^{d_{max}}\sum_{m_i=1}^{\min(d_i,m_{max})}\left(\frac{1}{2}
\right)^{d_i-m_i} & &
\\ \nonumber {N_c-d_i \choose m_i-\mu_i
}{ d_i-1 \choose m_i-1 }\mathrm{D}^{d_i}\mathrm{M}^{m_i}\mathrm{\Upsilon}^{\mu_i} & &
\\
\end{eqnarray}
where the $\mathrm{D}^{0} \mathrm{M}^{0}\mathrm{\Upsilon}^{0}=1$
monomial term corresponds to the case when there are no errors in
the specified subcode, and $\mu_i=\{0,1\}$ is the number of
boundary errors per subcode. As a result, the probability
enumerator for the entire interleave composed of $L_c$ EPCC
subcodes is given by $$\mathbf{\Lambda}^{L_c}(\mathrm{D},
\mathrm{M}, \mathrm{\Upsilon};\ m_{max}, d_{max})$$ given that
only $d_{max}$-weight error words $\mathbf{f}^{(i)}$ composed of
$m_{max}$ disjoint error patterns can occur per EPCC subcode,
where $d_{max}$ and $m_{max}$ are unbounded from above if EPCC
correction is turned off. The advantage of this approach is that
polynomial multiplication, or the more general multinomial
multiplication, can be performed efficiently by symbolic
manipulators, such as Maple\texttrademark, speeding up the
evaluation of (\ref{eqIVB7}) and (\ref{eqIVB5}). Utilizing the
compact, and easy-to-compute, probability enumerator, we can now
express the bound on the bit error rate of the TE-EPCC as:
\begin{align}\label{eqIVC2}
\nonumber P_b & \leq \frac{1}{K}\sum_{d_E=d^{min}_E}^{\infty}
Q\left(\frac{d_E}{\sigma}\right)\sum_{d=d_{min}}^{N}\frac{\mathbf{A}^o(d)\mathbf{\overline{A}}^i(d)}{{ N \choose d }} \sum_{\mu=0}^{L_c}\\
\nonumber & \mathop {\sum_{m=1}^{d}}\limits_{ m:\ \lambda_{cr}\geq
0, \lambda_{cr} \in \mathbf{N} } { d-m \choose \lambda_{cr} }
[\mathbf{\Lambda}^{L_c}(\mathrm{D},
\mathrm{M}, \mathrm{\Upsilon};\ \infty, \infty)]_{d,m,\mu} \\
\nonumber & -\mathop {\sum_{m=1}^{d}}\limits_{m:\ \lambda_{cr}=0
}[\mathbf{\Lambda}^{L_c}(\mathrm{D}, \mathrm{M},
\mathrm{\Upsilon};\ m_{c}, d_{c})]_{d,m,\mu} \\
\lambda_{cr} & = \frac{d_E^2-2 \alpha m+ \mu \alpha^2 -
(1-\alpha)^2 d}{4\alpha}
\end{align}
where the probability enumerator for a correctable EPCC codeword
is approximated by $$\mathbf{\Lambda}^{L_c}(\mathrm{D},
\mathrm{M}, \mathrm{\Upsilon};\ m_{c}, d_{c}),$$ for an EPCC of
maximum correction power $m_c$ per subcode, and $\mathbf{N}$ is
the set of natural numbers.
\begin{center}
\begin{table*} [!ht] 
    \caption{Interleaver gain exponent of the conventional non-precoded TE vs the TE-EPCC, $d_E^2=\{1, 2, 3, 4\}$.}
    \label{Table:IntlG1}
    \centering
    \begin{tabular}{|c|l|c|c|} \hline
        $d_E^2=1$ & \textrm{Error pattern classes}  & \textrm{TE} & \textrm{TE-EPCC} \\  \hline
        $\begin{array}{l} m=1 \\ \mu=1 \\ d=2 \rightarrow d_T \end{array}$  & \includegraphics*[width=1.5in]{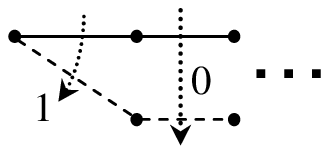}  & $\begin{array}{c} N^{-2} \\ \left( 1 \right) \end{array}$  & $\begin{array}{c} N^{-11} \\ \left( \frac{155925}{4} \right) \end{array}$  \\  \hline\hline
        $d_E^2=2$ & \textrm{Error pattern classes}  & \textrm{TE} & \textrm{TE-EPCC} \\  \hline
        $\begin{array}{l} m=1 \\ \mu=0 \\ d=2 \rightarrow d_T \end{array}$  & \includegraphics*[width=1.5in]{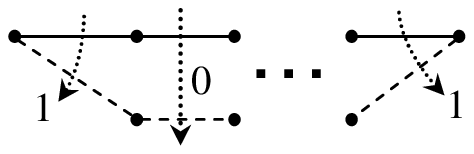}& $\begin{array}{c} N^{-1} \\ \left( 1 \right) \end{array}$  & $\begin{array}{c} N^{-10} \\ \left( \frac{155925}{4} \right) \end{array}$  \\  \hline\hline
        $d_E^2=3$ & \textrm{Error pattern classes}  & \textrm{TE} & \textrm{TE-EPCC} \\  \hline
        $\begin{array}{l} m=2 \\ \mu=1 \\ d=2 \rightarrow d_T \end{array}$  & \includegraphics*[width=2.5in]{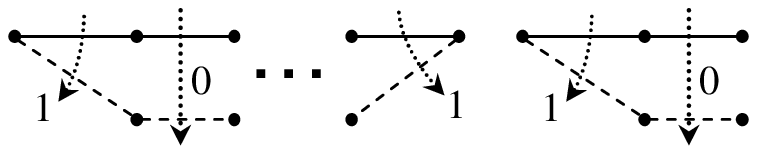}  & $\begin{array}{c} N^{-1} \\ \left( 2 \right) \end{array}$  & $\begin{array}{c} N^{-10} \\ \left( 779625 \right) \end{array}$  \\  \hline\hline
        $d_E^2=4$ & \textrm{Error pattern classes}  & \textrm{TE} & \textrm{TE-EPCC} \\  \hline
        $\begin{array}{l} m=2 \\ \mu=0 \\ d=2 \rightarrow d_T \end{array}$  & \includegraphics*[width=3in]{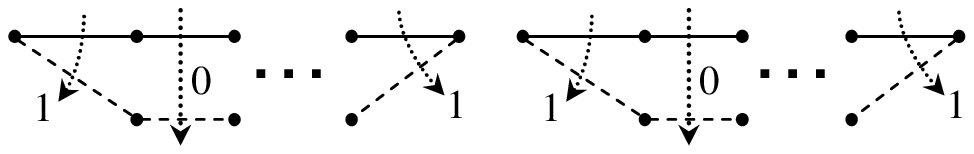}  & $\begin{array}{c} N^{0} \\ \left( 1 \right) \end{array}$  & $\begin{array}{c} N^{-9} \\ \left( \frac{779625}{2} \right) \end{array}$  \\  \hline
    \end{tabular}\\
\end{table*}
\end{center}

\begin{center}
\begin{table*} [!ht] 
    \caption{Interleaver gain exponent of the conventional non-precoded TE vs the TE-EPCC, $d_E^2=\{5, 6\}$.}
    \label{Table:IntlG2}
    \centering
    \begin{tabular}{|c|l|c|c|} \hline
        $d_E^2=5$ & \textrm{Error pattern classes}  & \textrm{TE} & \textrm{TE-EPCC} \\  \hline
        $\begin{array}{l} m=3 \\ \mu=1 \\ d=3 \rightarrow d_T \end{array}$  & \includegraphics*[width=3in]{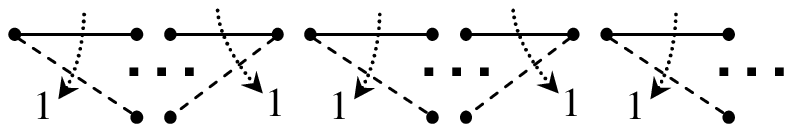}  & $\begin{array}{c} N^{-1} \\ \left( 3 \right) \end{array}$  & $\begin{array}{c} N^{-9} \\ \left( 3508313 \right) \end{array}$  \\  \hline
        $d_E^2=5$ & \textrm{Error pattern classes}  & \textrm{TE} & \textrm{TE-EPCC} \\  \hline
        $\begin{array}{l} m=1 \\ \mu=1 \\ d=2 \rightarrow d_T \end{array}$  & \includegraphics*[width=1.5in]{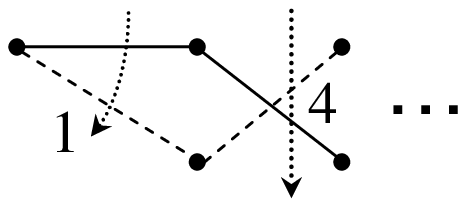}& $\begin{array}{c} N^{-2} \\ \left( 1 \right) \end{array}$  & $\begin{array}{c} N^{-2} \\ \left( 1 \right) \end{array}$  \\  \hline\hline
        $d_E^2=6$ & \textrm{Error pattern classes}  & \textrm{TE} & \textrm{TE-EPCC} \\  \hline
        $\begin{array}{l} m=3 \\ \mu=0 \\ d=3 \rightarrow d_T \end{array}$  & \includegraphics*[width=3in]{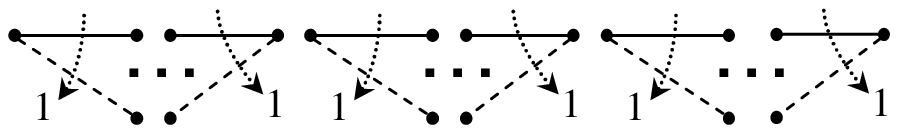}  & $\begin{array}{c} N^{0} \\  \left( 1 \right) \end{array}$  &  $\begin{array}{c} N^{-8} \\ \left( 1169438 \right) \end{array}$  \\  \hline
        $d_E^2=6$ & \textrm{Error pattern classes}  & \textrm{TE} & \textrm{TE-EPCC} \\  \hline
        $\begin{array}{l} m=1 \\ \mu=0 \\ d=2 \rightarrow d_T \end{array}$  & \includegraphics*[width=2in]{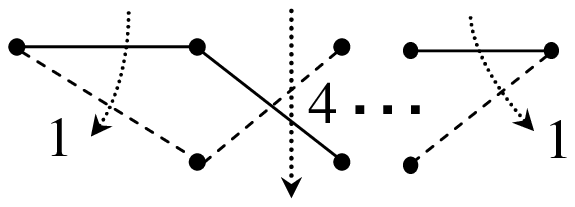}  & $\begin{array}{c} N^{-1} \\ \left( 1 \right) \end{array}$  & $\begin{array}{c} N^{-1} \\ \left( 1 \right) \end{array}$  \\  \hline
    \end{tabular}\\
\end{table*}
\end{center}

\begin{center}
\begin{table*} [!ht] 
    \caption{Interleaver gain exponent of the conventional non-precoded TE vs the TE-EPCC, $d_E^2=7$.}
    \label{Table:IntlG3}
    \centering
    \begin{tabular}{|c|l|c|c|} \hline
        $d_E^2=7$ & \textrm{Error pattern classes}  & \textrm{TE} & \textrm{TE-EPCC} \\  \hline
        $\begin{array}{l} m=4 \\ \mu=1 \\ d=4 \rightarrow d_T \end{array}$  & \includegraphics*[width=3in]{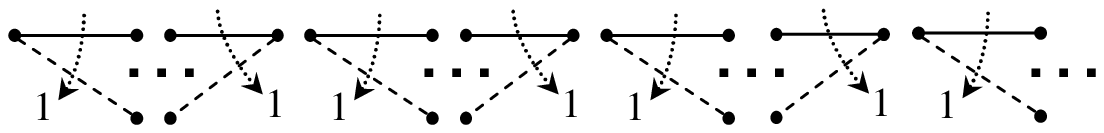}  & $\begin{array}{c} N^{-1} \\ \left( 4 \right) \end{array}$  & $\begin{array}{c} \left[N^{-1}\right]_{m_c=3} \\ \left( 4 \right) \\ \left[N^{-8}\right]_{m_c=4} \\ \left( 6237000 \right) \end{array}$  \\  \hline
        $d_E^2=7$ & \textrm{Error pattern classes}  & \textrm{TE} & \textrm{TE-EPCC} \\  \hline
        $\begin{array}{l} m=2 \\ \mu=1 \\ d=3 \rightarrow d_T \end{array}$  & \includegraphics*[width=2.5in]{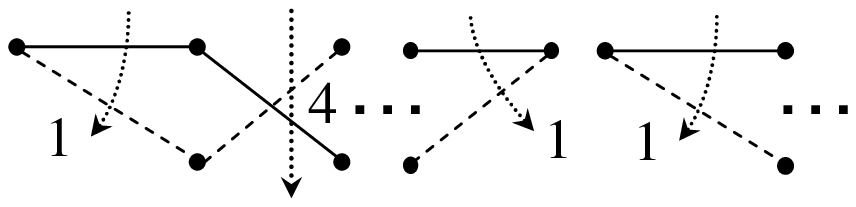}  & $\begin{array}{c} N^{-2} \\ \left( 6 \right) \end{array}$  & $\begin{array}{c} N^{-2} \\ \left( 6 \right) \end{array}$  \\  \hline
        $d_E^2=7$ & \textrm{Error pattern classes}  & \textrm{TE} & \textrm{TE-EPCC} \\  \hline
        $\begin{array}{l} m=2 \\ \mu=1 \\ d=3 \rightarrow d_T \end{array}$  & \includegraphics*[width=2.5in]{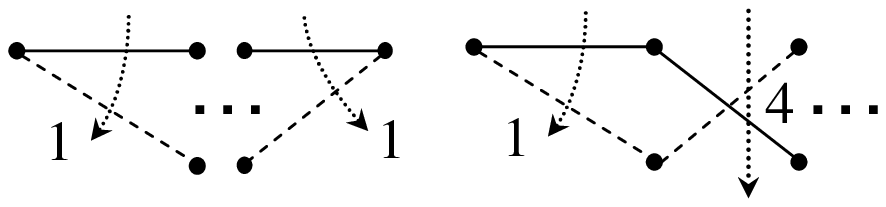}  & $\begin{array}{c} N^{-2} \\ \left( 6 \right) \end{array}$  & $\begin{array}{c} N^{-2} \\ \left( 6 \right) \end{array}$  \\  \hline
    \end{tabular}\\
\end{table*}
\end{center}

\begin{center}
\begin{table*} [!ht] 
    \caption{Interleaver gain exponent of the conventional precoded TE vs the TE-EPCC, $d_E^2=\{2, 3, 4, 5\}$.}
    \label{Table:IntlG4}
    \centering
    \begin{tabular}{|c|l|c|c|} \hline
        $d_E^2=2$ & \textrm{Error pattern classes}  & \textrm{precoded TE} & \textrm{TE-EPCC} \\  \hline
        $\begin{array}{l} m=1 \\ \mu=0 \\ d^H=2 \\ L = 2 \end{array}$  & \includegraphics*[width=2.0in]{dE2.eps}  & $\begin{array}{c} N^{-1} \\ \left( 2 \right) \end{array}$  & $\begin{array}{c} N^{-10} \\ \left( \frac{155925}{4} \right) \end{array}$  \\  \hline\hline
        $d_E^2=3$ & \textrm{Error pattern classes}  & \textrm{precoded TE} & \textrm{TE-EPCC} \\  \hline
        $\begin{array}{l} m=2 \\ \mu=1 \\ d^H=3 \\ L = 3 \end{array}$  & \includegraphics*[width=2.5in]{dE3.eps}& $\begin{array}{c} N^{-2} \\ \left( 6 \right) \end{array}$  & $\begin{array}{c} N^{-10} \\ \left( 779625 \right) \end{array}$  \\  \hline\hline
        $d_E^2=4$ & \textrm{Error pattern classes}  & \textrm{precoded TE} & \textrm{TE-EPCC} \\  \hline
        $\begin{array}{l} m=2 \\ \mu=0 \\ d^H=4 \\ L = 4 \end{array}$  & \includegraphics*[width=3in]{dE4.eps}  & $\begin{array}{c} N^{-2} \\ \left( 12 \right) \end{array}$  & $\begin{array}{c} N^{-9} \\ \left( \frac{779625}{2} \right) \end{array}$  \\  \hline\hline
        $d_E^2=5$ & \textrm{Error pattern classes}  & \textrm{precoded TE} & \textrm{TE-EPCC} \\  \hline
        $\begin{array}{l} m=3 \\ \mu=1 \\ d^H=5 \\ L = 5 \end{array}$  & \includegraphics*[width=3in]{dE5_1.eps}  & $\begin{array}{c} N^{-3} \\ \left( 60 \right) \end{array}$  & $\begin{array}{c} N^{-2} \\ \left( 1 \right) \end{array}$  \\  \hline
    \end{tabular}\\
\end{table*}
\end{center}

\section{Interleaver Gain Exponent of TE-EPCC}

To gain insight into how EPCC enhances TE performance, we pursue
an analytic approach to study the mechanism by which EPCC reduces
the multiplicity of low Euclidean distance errors. For this,
we limit our investigation to the dicode channel, for
which the spectrum of the Euclidean distance is comprised only of
integer values of $d_E^2$, and hence there are a fewer values that
$d_E$ can take in the lower range of the spectrum. The error probability shown in
 (\ref{eqIV3}) can be lowered by 1) increasing the minimum Euclidean
distance between error words, a traditional approach, or 2)
reducing the multiplicity of low Euclidean distance
errors, as in the turbo coding paradigm pioneered by
Berrou \textit{et al.}~\cite{Berrou93}. In turbo coding, the
coefficients of the error function for low Euclidean distances are
an inverse function of the interleaver size, $N$. For this reason,
turbo coding gain is often referred to as interleaver gain. At a
more detailed level, for the frequency of low weight errors to
asymptotically approach zero as the interleave size tends to
infinity, the exponent of the interleaver size in the
corresponding error coefficients should be less than zero.
Therefore, we can argue for the advantage of incorporating an EPCC in
TE, by showing how it works to decrease the exponent of
$N^{\delta}$ well below zero, especially for low Euclidean
distance errors. We call $\delta$ the interleaver gain exponent.
First, we isolate the exponent of $N$ in the expression of BER for
TE and TE-EPCC. The BER expression of the conventional TE (EPCC turned
off) for $\alpha=1$ is:
\begin{eqnarray}\label{eqV1}
\nonumber P_b & \leq & \frac{1}{K}\sum_{d_E=1}^{\infty}
Q\left(\frac{d_E}{\sigma}\right)\sum_{d=2}^{d_{T}}
 \frac{\mathbf{A}^o(d)\mathbf{\overline{A}}^i(d)}{{ N \choose d }} \\
\nonumber & & \sum_{\mu=0}^{1}\mathop {\sum_{m=1}^{d}}\limits_{
m:\ d_E^2-2m+\mu=0\ \textrm{mod}\ 4} { d-m \choose
\frac{d_E^2-2m+\mu}{4} } \\
\nonumber & & {\left(\frac{1}{2}\right)}^{d-m}{ N-d
\choose m-\mu } { d-1 \choose m-1 }\\
\end{eqnarray}
where $d_T \ll N$ is the truncated maximum error weight. We
truncated the Hamming error weight $d$ since large weight errors
correspond to larger Euclidean distances which have little contribution to the BER. To produce an expression for the
upper bound on BER with isolated powers of $N$, and at the same
time preserve it as an upper bound, we replace the binomial in the
denominator by the lower bound \cite{Benedetto98}:
$$ { N \choose d } > \frac{\left(N-d+1\right)^d}{d!}\simeq\frac{N^d}{d!}.$$
Moreover, to replace the binomial in the numerator with an upper
bound that is also a power of $N-d+1$, we first express it as:
$$
{ N-d \choose m-\mu }=\frac{m-\mu+1}{N-d+1}{ N-d+1 \choose m-\mu+1
}
$$
and employ the upper bound \cite{Benedetto98}:
$$
{ N-d+1 \choose m-\mu+1 } <
\frac{\left(N-d+1\right)^{m-\mu+1}}{(m-\mu+1)!} \simeq
\frac{N^{m-\mu+1}}{(m-\mu+1)!}.
$$
These bounds are tight when $N$ is large, and $d,m \ll N$, which
holds true in our case. Also we can upper bound the $Q$ function
by:
$$Q\left(\frac{d_E}{\sigma}\right) \leq \frac{1}{2}e^{-\frac{d^2_E}{2\sigma^2}}.$$
Substituting these approximate bounds in the BER upper bound in
(\ref{eqV1}), we get a looser but insightful bound:
\begin{eqnarray}\label{eqV2}
\nonumber P_b & < &
\frac{1}{2K}\sum_{d_E=1}^{\infty}\sum_{d=2}^{d_{T}}\sum_{\mu=0}^{1}
\mathop {\sum_{m=1}^{d}}\limits_{ m:\ d_E^2-2m+\mu=0\
\textrm{mod}\ 4} \\ & & \mathbf{B}_{d_E,d,m,\mu}
N^{m-\mu-d}e^{-\frac{d^2_E}{2\sigma^2}}
\end{eqnarray}
where $\mathbf{B}_{d_E,d,m,\mu}$ is given by:
\begin{eqnarray}\label{eqV3}
\nonumber \mathbf{B}_{d_E,d,m,\mu} & = &
\mathbf{A}^o(d)\mathbf{\overline{A}}^i(d)\frac{d!}{(m-\mu)!}
\\ & & {\left(\frac{1}{2}\right)}^{d-m} { d-m \choose
\frac{d_E^2-2m+\mu}{4} } { d-1 \choose m-1 }
\end{eqnarray}
For the sake of mathematical tractability, we study the interleaver gain exponent of $L_c=1$
TE-EPCC, i.e. single EPCC subcode per interleave. Utilizing the
same approximations as above in the BER bound of TE-EPCC for $L_c=1$
we get the expression:
\begin{eqnarray}\label{eqV4}
\nonumber P_b & < &
\frac{1}{2K}\sum_{d_E=1}^{\infty}e^{-\frac{d^2_E}{2\sigma^2}}\sum_{\mu=0}^{1}
\\
\nonumber & & \left[ \sum_{d=2}^{d_{T}} \mathop
{\sum_{m=1}^{d}}\limits_{ m:\ d_E^2-2m+\mu=0\ \textrm{mod}\
4}\mathbf{B}_{d_E,d,m,\mu}N^{m-\mu-d} \right. \\ \nonumber & &
\left. -\sum_{d=2}^{\min(d_{T},d_c)} \mathop
{\sum_{m=1}^{\min(d,m_c)}}\limits_{ m:\
d_E^2=2m-\mu}\mathbf{B}_{d_E,d,m,\mu}N^{m-\mu-d} \right].\\
\end{eqnarray}
The expression in (\ref{eqV4}) is just the expression in
(\ref{eqV2}) with those terms that are correctable by EPCC
subtracted. By identifying the maximum exponent of the interleaver
length $N$ in (\ref{eqV4}) and (\ref{eqV2}), we can compare the
asymptotic BER of TE and TE-EPCC in the limit of large interleaver
size. Assuming the minimum Hamming weight of the outer RSCC code
is $2$, we list the maximum interleaver gain exponent per $d^2_E$,
for TE and TE-EPCC ($d_c=10$, $m_c=\{3,4\}$, $L_c=1$) in
Table~\ref{Table:IntlG1} for $d^2_E=\{1, \ldots, 4\}$, in
Table~\ref{Table:IntlG2} for $d^2_E=\{5, 6\}$, and in
Table~\ref{Table:IntlG3} for $d^2_E=7$. We also list for each
$d^2_E$, the generating error patterns and their corresponding
parameters $d$, $m$, and $\mu$. In addition, under each
interleaver gain exponent, we list in parenthesis the
corresponding multiplicative coefficient
$\mathbf{B}_{d_E,d,m,\mu}$, excluding the term
$\mathbf{A}^o(d)\mathbf{\overline{A}}^i(d)$ relating to the outer
RSCC Hamming error weight distribution.

For the precoded TE, using the same approximations as above, it can be
shown that the BER bound of Appendix A is dominated by the terms:
\begin{eqnarray}\label{eqV5}
P_b & < & \frac{1}{2K}\sum_{d_E=2}^{\infty}\sum_{d=2}^{N}
\mathbf{B}_{d} N^{-\lceil \frac{d}{2} \rceil}
e^{-\frac{d^2_E}{2\sigma^2}}
\end{eqnarray}
where $\mathbf{B}_{d}$ is given by:
\begin{eqnarray}\label{eqV6}
\mathbf{B}_{d}= \mathbf{A}^o(d)\mathbf{\overline{A}}^i(d) \lceil
\frac{d}{2} \rceil { d \choose \lceil \frac{d}{2} \rceil }.
\end{eqnarray}
In the derivation of the above bound we only kept terms whose
error length $L$ is equal to the Hamming distance $d^H$ since
they have the dominant interleaver gain exponent at each $d^H$.
In this case, the dominant error will have $d_E^2 = d^H$ as shown
in Table~\ref{Table:IntlG4}.

First, we note that for the non-precoded TE, the interleaver gain
exponents are all negative for $d^2_E=1$ to $d^2_E=3$, which are
the terms that dominate the BER for medium to high SNRs. Second, we
note that the error patterns, for this same range of error
Euclidean distances, up to $d=10$, all belong to the dominant error
class. As a result, the TE-EPCC manages to substantially decrease the
interleaver gain exponent by a factor of $N^{9}$. Also, for
$d^2_E=4$, where the TE does not achieve any interleaver gain, the TE-EPCC
has an impressive interleaver gain exponent of $N^{-9}$.

The extremely low exponents suggest that the TE-EPCC will have large
gain even for relatively short interleavers, and would thus
deliver satisfactory gain for short to medium RSCC codeword sizes.
At the same time, for such short interleavers, the TE would
considerably suffer in terms of turbo gain. These conclusions will
be numerically demonstrated in the next section by evaluating the
BER bound for interleavers as short as $100$ bits. Furthermore,
although $\mathbf{B}_{d_E,d,m,\mu}$ is significantly larger in
the TE-EPCC compared to the TE for the same $d^2_E$, the term
$\mathbf{B}_{d_E,d,m,\mu}N^{m-\mu-d}$ is still several orders of
magnitude lower for the TE-EPCC.

Although less important, we also show the interleaver gain for
higher error Euclidean distances in Table~\ref{Table:IntlG2} and
Table~\ref{Table:IntlG3}. Most notably, the TE-EPCC ($d_c=10$, $m_c=3$,
$L_c=1$) corrects errors belonging to the dominant error class for
$d^2_E=5$ and $d^2_E=6$, lowering, in the process, the maximum
interleaver gain exponent by a factor of $N$, a turbo gain
that becomes more substantial for large interleavers. Actually,
for $d^2_E=6$, the TE possess no interleave gain, while TE-EPCC BER is
dominated by the non-targeted set of non-dominant errors that
result in the exponent $N^{-1}$, still achieving an interleaver gain.
On the other hand, the TE-EPCC ($d_c=10$, $m_c=3$, $L_c=1$)
would offer no advantage when $d^2_E=7$. Note that although all
errors belong to $\chi^{dom}$ when $m=4$, their multiplicity $m$
exceeds the maximum multiple-error-pattern correction capability
of $m_c=3$. However, the TE-EPCC ($d_c=10$, $m_c=4$, $L_c=1$) manages
to reduce the maximum interleaver gain exponent to $N^{-2}$, by
reducing the contribution of $\chi^{dom}$ to $d^2_E=7$ by a factor
of $N^{7}$.

Comparing the interleaver gain exponents of the TE-EPCC and the
precoded TE in Table~\ref{Table:IntlG4}, we note that the TE-EPCC
focuses on error events of $d_E^2 \leq 4$ and $d^H \leq 4$,
reducing the interleaver exponent by a factor of $N^9$, $N^8$, and
$N^7$ at $d_E^2=\{2,3,4\}$, respectively, compared to the precoded
TE. However, the precoded TE's interleaver exponent is lower by a
factor of $N$ compared to the TE-EPCC at $d_E^2=5$. Hence, we
predict that the TE-EPCC's BER floor will be far lower than that
of the precoded TE, while the precoded TE waterfall BER may still
be lower owing to the lower interleaver gain exponents of higher
Euclidean-distance error events. In summary, the EPCC shapes the
error weight spectrum to improve the error floor while preserving
gains achieved in the waterfall region. The TE-EPCC is thus a
novel turbo equalization approach that enhances the spectrum
thinning at low weights, where the error floor is not a strong
function of interleaver size as the waterfall region is. Actually,
all ML type bounds, including our bound, study the error floor
rather than the waterfall region. This works in our favor since
our EPCC code's advantage lies there. In addition, our EPCC works
in a probabilistic fashion to enhance the minimum Euclidean
distance, in addition to the interleaver gain, compared to trellis
constrained methods that directly increase the minimum Euclidean
distance.

\section{A SISO Decoder for EPCC}\label{practicaldec}
We have thus far presented the ideal behavior expected from a
perfect EPCC decoder. We now discuss a practical implementation of
EPCC SISO decoding based on the algebraic single-pattern
correcting decoder of Section \textrm{II} and the soft side
information made available by the channel observations
$\mathbf{r}$ and the outer RSCC SISO decoder.

In the decoder flow chart, a decision is first made on whether the
hard input of the decoder $\mathbf{\hat{c}}$ contains a single or
a multiple error pattern via the syndrome check. If the initial
syndrome check indicates either an error free input, or else, a
single error pattern with high reliability~\cite{Jih_Moon}, then
the following formula is used to generate the soft decision
reliability for the $k$-th hard bit in the corrected codeword
\cite{Pyndiah98}:
\begin{eqnarray}\label{eqVI1}
  \lambda_k=\beta^{iter} \times \lambda_{max} \times \hat{d}_{k}
\end{eqnarray}
where $\mathbf{\hat{d}}$ is the bipolar representation of the
error-free/corrected bit, $\lambda_{max}$ is a preset value for
the maximum reliability at convergence of turbo performance, and
the multiplier $\beta^{iter} < 1$ is useful in incorporating the
EPCC SISO decoder in the iterative loop. Note that in an iterative
system the level of confidence in bit decisions is lower at the
initial iterations, and thus multiplying the generated log
likelihood ratios by the back-off factor $\beta^{iter}$ reduces
the risk of error propagation. On the other hand, if the computed
syndrome is unrecognized, signaling a multiple occurrence, then
the list decoder is activated which involves computing
correlator-based reliability estimates for local dominant patterns
in the ML word, as will be explained later in this section. Simulations
show that the aforementioned strategy of moving between
list-decoding and algebraic single pattern decoding results in
improved performance compared to running list-decoding all the
time, since at later turbo iterations single error-pattern
occurrences are more likely, and syndrome-decoding is more robust
in such scenarios.

In communicating with the other building blocks of the turbo
system, the EPCC decoder receives the interleaved extrinsic LLR $\lambda$
coming from the outer RSCC code as an \emph{a priori} input in
calculating its error pattern \emph{a posteriori} probabilities.
On the other hand, in the final soft output stage, after
generating a list of the most probable candidate codewords and
their likelihoods, the decoder uses the list to calculate the
output bit-level decision reliabilities that serve as the \emph{a
priori} input LLR to the outer RSCC SISO decoder.

\begin{figure*}[hbtp]
\centering
\includegraphics*[width= 6 in]{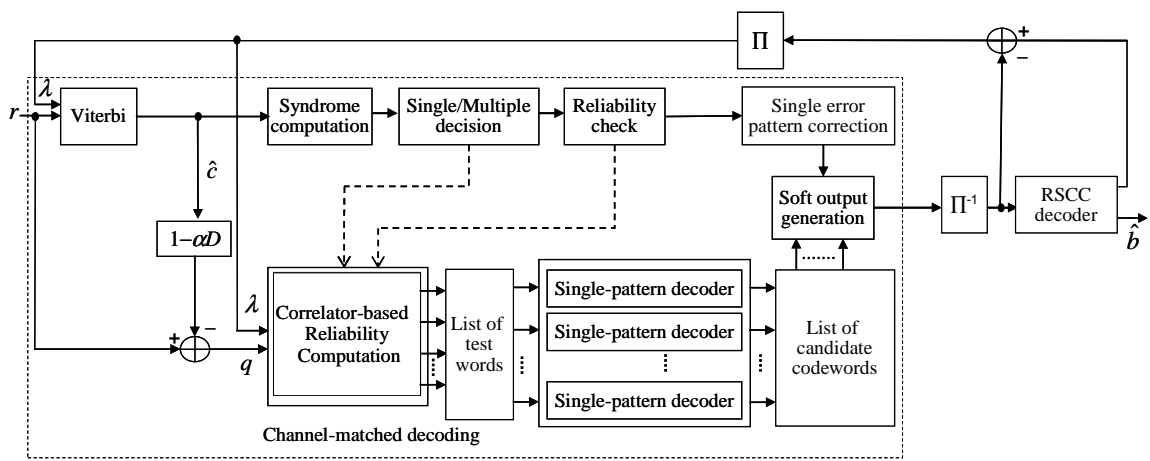}
\caption{The TE-EPCC block diagram.}\label{Turbo_dec}
\end{figure*}
The internal workings of the EPCC soft output list decoder
considered here consists of four stages, see Fig.~\ref{Turbo_dec}:
\begin{itemize}
    \item The probability of a dominant single error event is estimated at each likely starting position.
    \item The test error word list is generated by inserting the most probable combination of dominant error patterns into the channel detector ML output.
    \item An array of parallel single-pattern correcting decoders decode the test words to produce a list of valid codewords and accompanying likelihoods.
    \item The list of candidate codeword likelihoods is used to generate bit-level decisions along with their reliabilities.
\end{itemize}

\subsection{Dominant Error Probability Estimation}
At this decoder stage we estimate the likelihood $C(e_k^{(i)})$ at
all possible locations $k$ that the dominant error event $e^{(i)}$
can occur. A bank of error event correlators can be used for this
purpose as was shown in~\cite{Jih_Moon} and ~\cite{Hakim07}. Let
$r_k$ be the channel detector input sequence $r_k = {c}_k * h_k +
w_k$, where ${c}_k$ is the bipolar representation of the
transmitted codeword sequence, $h_k$ is the channel response of
length $l_h$, and $w_k$ is zero-mean AWGN noise with variance
$\sigma^2$. Also, let ${q}_k = r_k - (\hat{c}_k * h_k) = (c_k -
\hat{c}_k)*h_k + w_k $ be the channel detector's output error
sequence. If a target error pattern sequence $e_k^{(i)}$ occurs at
positions from $k = j$ to $k = j+l_i-1$, then $q_k$ can be written
as
 \begin{align}\label{eqVI2}
 \begin{split}
 q_k &= [\textbf{c}- \hat{\textbf{c}}^{(i)}]_j^{j+l_i-1}*h_k + w_k\\
     &= [\textbf{e}^{(i)}]_j^{j+l_i-1}*h_k + w_k \\
     &= [\mathbf{\xi}^{(i)}]_j^{j+l_i^h} + w_k
 \end{split}
 \end{align}
where $\xi_k^{(i)}$ is the channel response of the error sequence,
and is given by $\xi_k^{(i)} = e_k^{(i)}*h_k$, and $l_i^h = l_i +
l_h -2$. Using the MAP criterion, we can derive an estimate of the
likelihood of $e_k^{(i)}$ by measuring the distance of the
resulting $\mathbf{q}(e_k^{(i)})$ to the channel observation
$\mathbf{r}$ relative to the error free sequence
$\mathbf{q}(\hat{\textbf{c}})$, which simplifies to just the
difference in Euclidean distances of $\mathbf{q}(e_k^{(i)})$ and
$\mathbf{q}(\hat{\textbf{c}})$ measured to $\mathbf{r}$ in the ML
sense. The ML postulate becomes useful if $\hat{\textbf{c}}$ is
assumed i.i.d,
or when it is the best that can be done when no
\textit{a priori} side information is available.
In practice, though, the side
information in the form of bit-level log-likelihood ratios
$\lambda_k$ can be efficiently provided by the outer constituent
code in the turbo sense. For each $e_k^{(i)}$, we then
estimate~\cite{Jih_Moon}:
\begin{eqnarray}\label{eqVI3}
\nonumber C(e_j^{(i)}) & = & \sum\limits_{k = j}^{j + l_i^h }
\frac{1}{{2\sigma ^2 }}\left( {q_k ^2  - (q_k  - \xi_k^{(i)} )^2 }
\right) \\
& & - \left( \sum\limits_{k = j,\hat c_k =  + 1}^{j + l_i^h }
{\lambda _k }  - \sum\limits_{k = j,\hat c_k =  - 1}^{j + l_i^h }
{\lambda _k }\right)
\end{eqnarray}
where $\lambda_k$ is the \textit{a priori} LLR of the error-event
bit at position $k$ as received from the outer soft decoder, and
we are assuming here that error event sequences do not include $0$
bits, i.e., the ML sequence and error sequence do not agree for
the entire duration of the error event. Finally, the expanded list
of dominant errors and their likely positions is sorted according
to the computed reliabilities.

\subsection{Generation of the Test Error Word List}
In order to expand the decoding-sphere radius of the algebraic
single-error correcting code, and to benefit from channel side
information, we adopt a list-decoding structure that resembles
Chase decoding~\cite{Chase} in the sense of generating test
vectors at the parallel decoder input. However, a pivotal
difference in the methodologies is in the test word construction
stage of the EPCC decoder, where we flip multi-bit dominant error
patterns, rather than individual independent bits as in Chase
decoding. The resulting list of test vectors may contain one or
more words that are just one dominant error pattern away from the
correct codeword in terms of Hamming distance. Hence, if the
resulting set of test error words is decoded by an array of
single-pattern correcting decoders, then, one or more codewords in
the list of valid candidate codewords can be the correct codeword
with high probability. This novel pattern-level extension of the
Chase decoding algorithm was first proposed in~\cite{Hakim07} in
the context of SISO decoding of EPCC as a building block in TE,
and in~\cite{Jih_Moon} in the context of list decoding of
algebraic single-pattern correcting EPCC. Recently, a
pattern-flipping Chase was also studied in~\cite{kumar10}. This
later approach differs from the earlier work
in~\cite{Hakim07,Jih_Moon} only in the block that estimates
pattern reliabilities, where SOVA is utilized to estimate error
event probabilities instead of a bank of error-event-matched
correlators.

The probability measure of a given test word with a particular
combination of dominant error patterns is the product of the
probabilities of the constituent errors. In the construction of
test words, we select the most probable errors in the sense of
maximizing the correlator function of~(\ref{eqVI3}).

The requirement to have $m_c$-error-pattern-correction capability
using the single-pattern correcting decoders, dictates that test
words must include up to $m_c-1$ single dominant error patterns.
Starting from the $l_{max}$ most probable such dominant errors,
one can think of $\sum_{j=1}^{m_c-1}{l_{max} \choose j}$ ways of
corrupting the ML word with up to $m_c-1$ local error patterns.
From this large set of potential combinations, a relatively small
subset of most probable combinations needs to be chosen to
maintain reasonable complexity. One can think of many different
ways of effectively constructing such a list~\cite{Jih_Moon} based
on the probable local error patterns that have been identified.
\subsection{Parallel Algebraic Decoding}
The list of test error words generated above is delivered to an
array of single-error-pattern correcting decoders that work in
parallel to generate the candidate codeword list. The number of
parallel decoders is identical to the size of the test word list,
and is a crucial parameter that controls the EPCC decoder's
complexity/performance tradeoff.

\subsection{Generation of Soft Output}
The candidate codeword list constructed by our ``pattern-level"
list decoder is used to calculate the more familiar bit-level
reliabilities that constitute the output soft information supplied
by the EPCC SISO decoder. We measure the probability of a
candidate codeword given the observed word by the product of the
probabilities of each ``local" error pattern forming the candidate
word. Specifically, let $\mathbf{c}$ represent a candidate
codeword with, say, $K$ error-pattern corruption with respect to
the ML word $\mathbf{\hat{c}}$. Then, the \emph{a posteriori}
probability of this particular test word,
$Pr(\mathbf{c}|\mathbf{\hat{c}},\mathbf{r})$, is estimated by
multiplying the probability estimates of the $K$ local patterns,
given the channel observation $\mathbf{r}$ at the detector input.

Given the list of codewords and their accompanying \emph{a
posteriori} probabilities, the reliability $\lambda_k$ of the
coded bit $c_k$ is evaluated as:
\begin{eqnarray}\label{eqVI4}
  \lambda_k=\log \frac{\sum_{\mathbf{c} \in \textbf{S}_{k}^{+}}Pr(\mathbf{c}|\mathbf{\hat{c}},\mathbf{r})}{\sum_{\mathbf{c} \in
  \textbf{S}_{k}^{-}}Pr(\mathbf{c}|\mathbf{\hat{c}},\mathbf{r})}
\end{eqnarray}
where $\textbf{S}_{k}^{+}$ is the set of candidate codewords where
$c_k=+1$, and $\textbf{S}_{k}^{-}$ is the set of candidate
codewords where $c_k=-1$. The quantity in~(\ref{eqVI4}) is
utilized when the candidate codewords do not all agree on the bit
decision for location $k$. In the event that all codewords do
agree on the decision for $c_k$, a method used by \cite{Pyndiah98}
is adopted for generating soft information as follows
\begin{eqnarray}\label{eqVI5}
  \lambda_k=\beta^{iter} \times \lambda_{max} \times \hat{d}_{k}
\end{eqnarray}
where $\hat{d}_k$ is the bipolar representation of the agreed-upon
decision, and $\lambda_{max}$ and $\beta^{iter} < 1$ have the same
function as in~(\ref{eqVI1}).

\section{Numerical Analysis and Simulation Results}

Utilizing the analytic approximation of the BER of conventional TE
and TE-EPCC systems, we study the relative performance of these
systems for different levels of the severity of ISI. We also study the special
case of the dicode channel $1-D$ in a variety of channel
conditions. We will assume throughout the analysis that the SNR rate penalty
(in dB) is proportional to $10\log_{10}\frac{1}{R}$, where $R$ is the code rate.

\subsection{BER-Bound Validation for the Dicode Channel}
\begin{figure}[hbtp]
\centering
\includegraphics*[width= \figsz in]{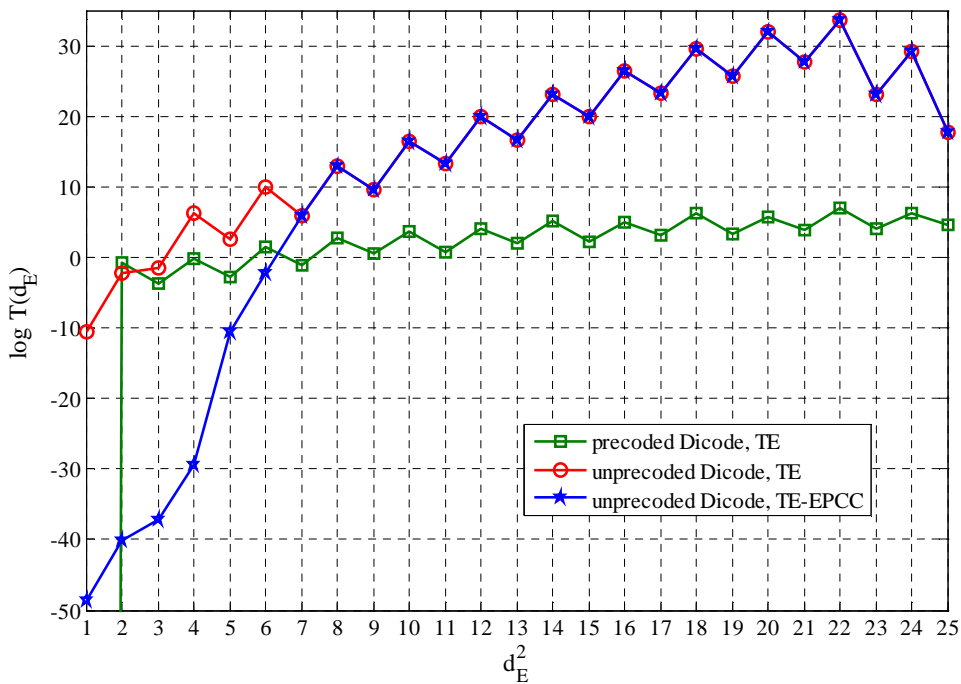}
\caption{$\log \overline{T}(d_E)$ for various TE systems, and
$(31,33)$ RSCC.}\label{Tde_comp}
\end{figure}

The log of the average Euclidean distance distribution of the
dicode channel, $\log \overline{T}(d_E)$, is shown in
Fig.~\ref{Tde_comp} for conventional TE and TE-EPCC systems.
Fig.~\ref{Tde_comp} also includes the Euclidean error distribution
for the precoded TE, derived in a similar way to that
of~\cite{Oberg01}. $\log \overline{T}(d_E)$ is calculated for a TE
with $K=4096$, a rate $1/2$ base constituent RSCC, punctured to
rate $R=8/9$ with generator polynomial connections $(31,33)$ in
octal format, and $L_c=7$ EPCC with $m_c=3$ and $d_c=10$. Each
EPCC subcode is a $(630,616)$ systematic cyclic code of rate
$0.98$ shortened to accommodate $L_c=7$ subcodes per interleave.
From the average Euclidean distance distribution, we can conclude
that the precoded TE exhibits larger interleaver gains compared to
the non-precoded TE in the waterfall region, i.e. low to medium
SNRs. This is because $\log \overline{T}(d_E)$ is lower for the
precoded TE everywhere when $d_E^2> 3$. However, for higher SNRs,
at the error floor region, the contribution of squared Euclidean
distance $2$ becomes stronger, and as seen in the figure, the
average number of Hamming weight $2$ errors that generate
$d^2_E=2$ is more for the precoded compared to the non-precoded
dicode channel. On the other hand, the EPCC concentrates on low
Euclidean distances, reducing their frequency substantially up to
$d^2_E=6$. This results in improved BER in the error floor and
yields a similar waterfall BER compared to the conventional
unprecoded TE.
\begin{figure}[hbtp]
\centering
\includegraphics*[width= \figsz in]{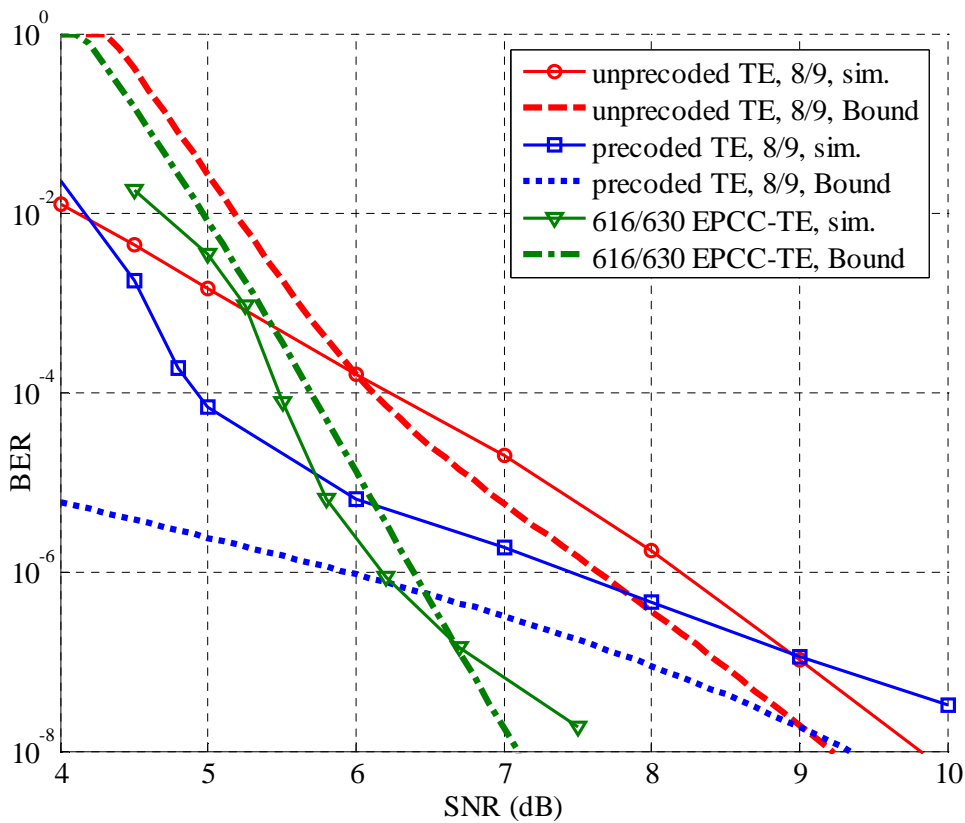}
\caption{Simulated vs bounded BER for various TE systems using
$(31,33)$ RSCC.}\label{N4608_EPCC}
\end{figure}

In Fig.~\ref{N4608_EPCC}, the simulated BER is shown for
conventional TEs with precoded and non-precoded dicode channels
and the TE-EPCC with a non-precoded dicode channel. Moreover, the simulated BER
is compared with the estimated BER bound computed for the same
parameters as in Fig.~\ref{Tde_comp}. The TE-EPCC is decoded via the
practical soft decoder described earlier in Section
\textrm{\ref{practicaldec}}, where we implement $5$ turbo
iterations of the non-precoded TE and $10$ turbo iterations of
precoded TE and TE-EPCC systems; we used up to $100$ test patterns in
the list decoder of the TE-EPCC. The number of turbo iterations is
chosen for each system such that the turbo
gain saturates. The figure shows that the TE-EPCC has definite performance
advantage in the low error rate region.
The actual simulation curve comes above the analytical bound for the TE-EPCC
at very low BERs. This arises from imperfect uniform interleaving in the
practical decoders as also pointed out in~\cite{Oberg01}. Nonetheless,
the actual gain gaps between the TE-EPCC and the conventional TEs seem even large than
predicted by the bound; this is mainly attributed to
the higher sensitivity of conventional TEs to the interleaver
design compared to the TE-EPCC, an argument based on the interleaver
gain exponent of both systems.

\begin{figure}[hbtp]
\centering
\includegraphics*[width= \figsz in]{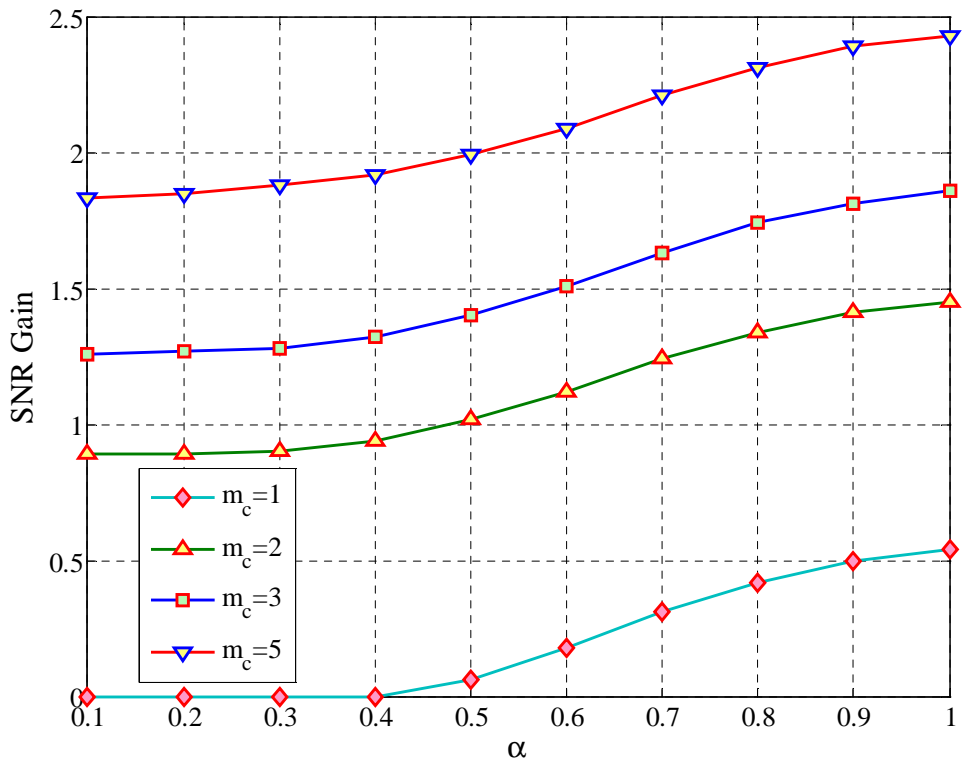}
\caption{SNR gain (dB) of the TE-EPCC over the non-precoded TE at
a BER of $10^{-7}$ as function of ISI severity level $\alpha$ for
outer RSCC $(7,5)$, punctured $R=8/9$, and interleaver size
$N=616$.}\label{diffalpha}
\end{figure}
\subsection{The Severity of ISI}

In Fig.~\ref{diffalpha} we plot the SNR gain of the TE-EPCC over
the non-precoded TE defined as the difference in the minimum SNR
required to achieve a BER of $10^{-7}$ for both systems. The BER
of the non-precoded TE improves as $\alpha \rightarrow 0$, since
the Euclidean distance of dominant Hamming errors grows with their
Hamming distance when $\alpha < 1$, where the error length is
linearly proportional to $(1-\alpha^2)>0$. On the other hand, for
a given EPCC correction power $m_c$, the BER of the TE-EPCC
remains almost the same as $\alpha \rightarrow 0$ since dominant
Hamming errors are correctable irrespective of their Euclidean
weight. The net result is a higher SNR gain furnished by the
TE-EPCC as $\alpha \rightarrow 1$. Furthermore, for a given
$\alpha$, the SNR gain of the TE-EPCC grows as $m_c$ is increased,
where a $2$ dB improvement can be achieved by increasing $m_c$
from $1$ to $5$ for all the range of $\alpha$.
\begin{figure}[hbtp]
\centering
\includegraphics*[width= \figsz in]{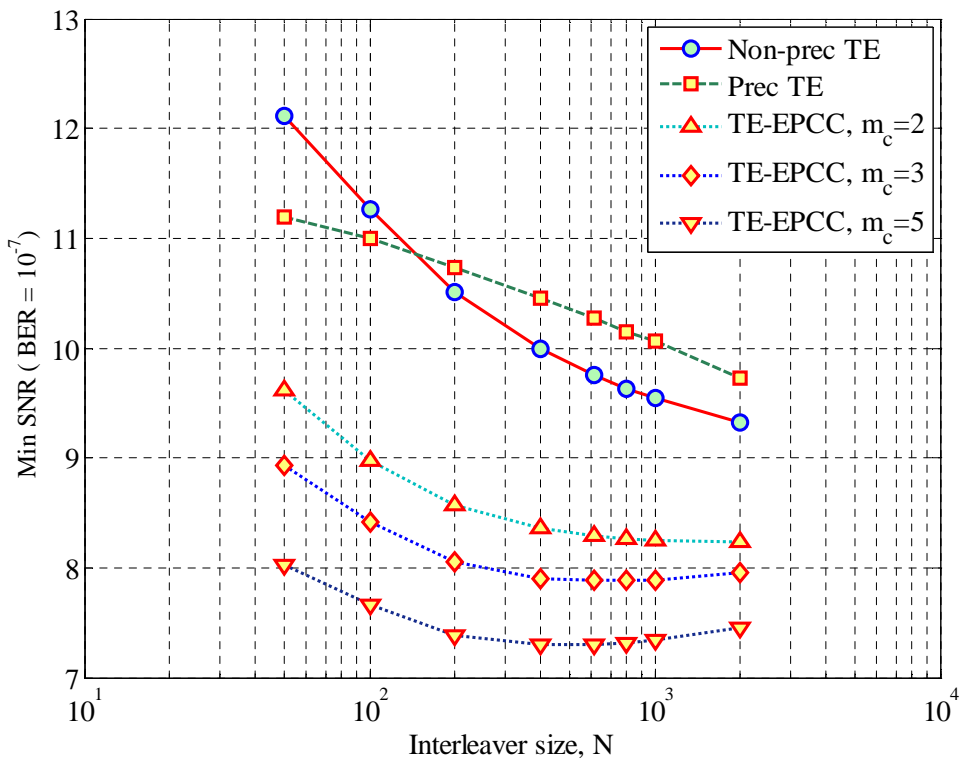}
\caption{Minimum SNR (dB) required to achieve a BER of $10^{-7}$
using $(7,5)$ RSCC, punctured rate $8/9$, TE-EPCC
$\{m_c=3,d_c=10,L_c=1\}$, and different interleaver sizes.}
\label{diffNg071en7}
\end{figure}
\subsection{Interleaver Gain}
In Fig.~\ref{diffNg071en7} we compare the minimum SNR to achieve a
BER of $10^{-7}$. The curves of the conventional TE with the
precoded and non-precoded dicode channels and the TE-EPCC for the
non-precoded dicode channel are shown for interleaver lengths
$N=\{50,100,200,400,600,800,1000,2000\}$ bits, punctured rate
$R=8/9$ RSCC with connections $(7,5)$ in octal format, and EPCC
with $m_c=\{2,3,5\}$ and $d_c=10$. The SNR gain of the TE-EPCC
over the non-precoded TE using $N=50$ is $2.3$ dB, $3$ dB, and $4$
dB for EPCC correction powers $m_c=\{2,3,5\}$, respectively. On
the other hand, using $N=2000$, this shrinks to $1.1$ dB, $1.3$
dB, and $1.8$ dB for EPCC correction powers $m_c=\{2,3,5\}$,
respectively.

We note that as the interleaver size $N$ of the TE-EPCC increases,
the turbo gain of TE-EPCC increases accordingly. Also, since we
maintain the same number of parity bits as the codeword length
increases, less SNR rate penalty is incurred as $N$ increases. On
the other hand, the probability of $m > 1$ multiple errors per
subcode increases for larger $N$, surpassing EPCC's correction
capability. Due to these conflicting effects of the TE-EPCC, its
minimum SNR plateaus and even increases as $N$ increase beyond a
certain point. All in all, the relative advantage of the TE-EPCC
in practical system seems most visible with small interleaver
sizes. Furthermore, as can be observed in the figure, increasing
$m_c$ is also most effective for smaller interleaver sizes.

In practical EPCC code construction, in order to obtain shorter
EPCC code lengths, while serially concatenating one EPCC subcode
per RSCC interleave, i.e. $L_c=1$, the EPCC code length is
shortened from the long $(630,616)$ EPCC at the same level of
redundancy. While to support interleaver sizes above $630$, we
duplicate EPCC subcodes, i.e. $L_c>1$, and use shortening to fit
fractions of EPCC subcodes in one interleave. For instance, we
implement $(114,100)$ EPCC of rate $0.88$ for interleaver length
$N=100$, and long $(630,616)$ EPCC + shortened $(398,384)$ EPCC
for $N=1000$.

\subsection{SNR Gain as Function of $L_c$ and $m_c$}

\begin{figure}[hbtp]
\centering
\includegraphics*[width= \figsz in]{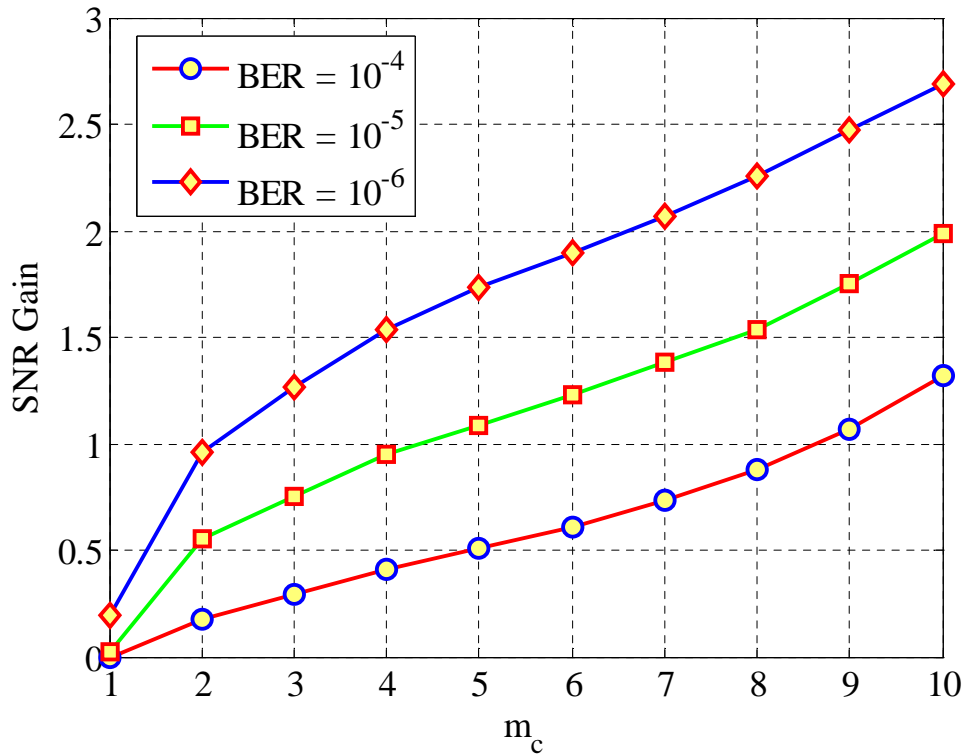}
\caption{SNR gain of the TE-EPCC over the non-precoded TE at
several BER operating points as function of $m_c$ for the dicode
channel, outer RSCC $(7,5)$, punctured $R=8/9$, and interleaver
size $N=616$.}\label{diffmc_N616}
\end{figure}

The performance of TE-EPCC can be further improved by increasing
its multiple error correction capability $m_c$, per subcode.
However, the complexity of the practical decoder would increase
accordingly as more test words have to be constructed in the list
decoder. Fig.~\ref{diffmc_N616} shows TE-EPCC's SNR gain over the
non-precoded TE for several BER operating points, $N=1200$,
punctured $R=8/9$ RSCC with connections $(7,5)$, and $L_c=1$ EPCC
with maximum correction capability increased from $m_c=1$ to
$m_c=10$ and $d_c=10$. The curves demonstrate that TE-EPCC's SNR
gain grows almost linearly as $m_c$ is increased.
\begin{figure}[hbtp]
\centering
\includegraphics*[width= \figsz in]{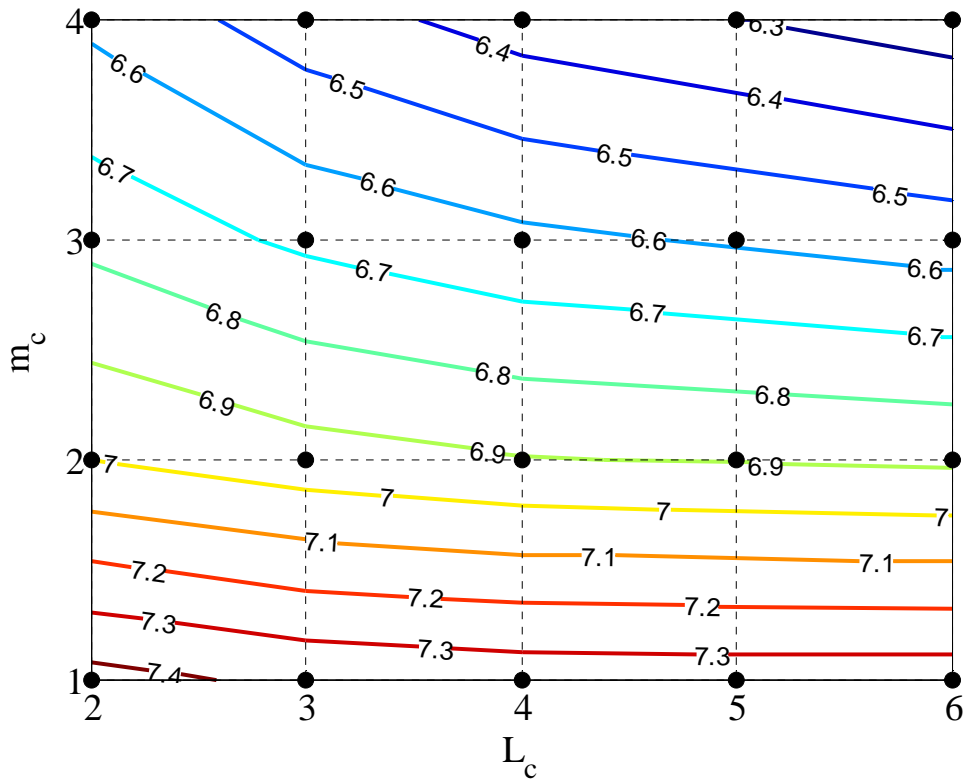}
\caption{An interpolated contour plot of the minimum SNR required
by TE-EPCC to achieve BER=$10^{-5}$ for different combinations of
$m_c$ and $L_c$, N=$1200$, and RSCC $(7,5)$ of punctured
$R=8/9$.}\label{diffL_diffM}
\end{figure}
Another design method to increase the correction capability of
TE-EPCC, without considerably increasing its complexity, is to use
$L_c>1$ EPCC subcodes per interleave. To study the design space
spanned by $m_c$ and $L_c$ for a given interleave, we evaluate the
BER bound for the set composed of the Cartesian product of the
sets $m_c=\{1,2,3,4\}$ and $L_c=\{2,3,4,5,6\}$. Then we plot a
continuous contour of the minimum SNR to achieve BER=$10^{-5}$ by
interpolating the values found at the elements of the Cartesian
product. A contour plot with an SNR step of $0.1$ dB is shown in
Fig.~\ref{diffL_diffM} for $N=1200$, punctured-rate $8/9$ RSCC
with connections $(7,5)$, and EPCC with different combinations of
$m_c$ and $L_c$, and $d_c=10$. We note that the combinations
$\{L_c=2,m_c=4\}$ and $\{L_c=5,m_c=3\}$ require a similar minimum
SNR=$6.6$ dB to achieve a floor BER of $10^{-5}$. Nonetheless, as
can be seen in Fig.~\ref{diffL_diffM} the slope the equi-SNR
contour lines decreases for higher $L_c$ and lower $m_c$. This
means that as the number of subcodes $L_c$ increases per
interleave, the correction capability plateaus, especially when
$m_c \leq 2$. This is due to the higher level of redundancy
required for shortened EPCC to maintain the maximum correction
capability $m_c$ of longer EPCC. For instance, at one extreme, to
maintain the correction capability at a shortened EPCC code length
of $44$ bits, i.e. $L_c=40$ and $N=1200$, a shortened EPCC of rate
$0.68$ would incur a staggering rate penalty of $1.7$ dB. An
alternative concatenation approach that avoids the rate penalty of
serial concatenation to a short inner EPCC is discussed
in~\cite{Hakim10}.

\subsection{Puncturing Rate}

\begin{figure}[hbtp]
\centering
\includegraphics*[width= \figsz in]{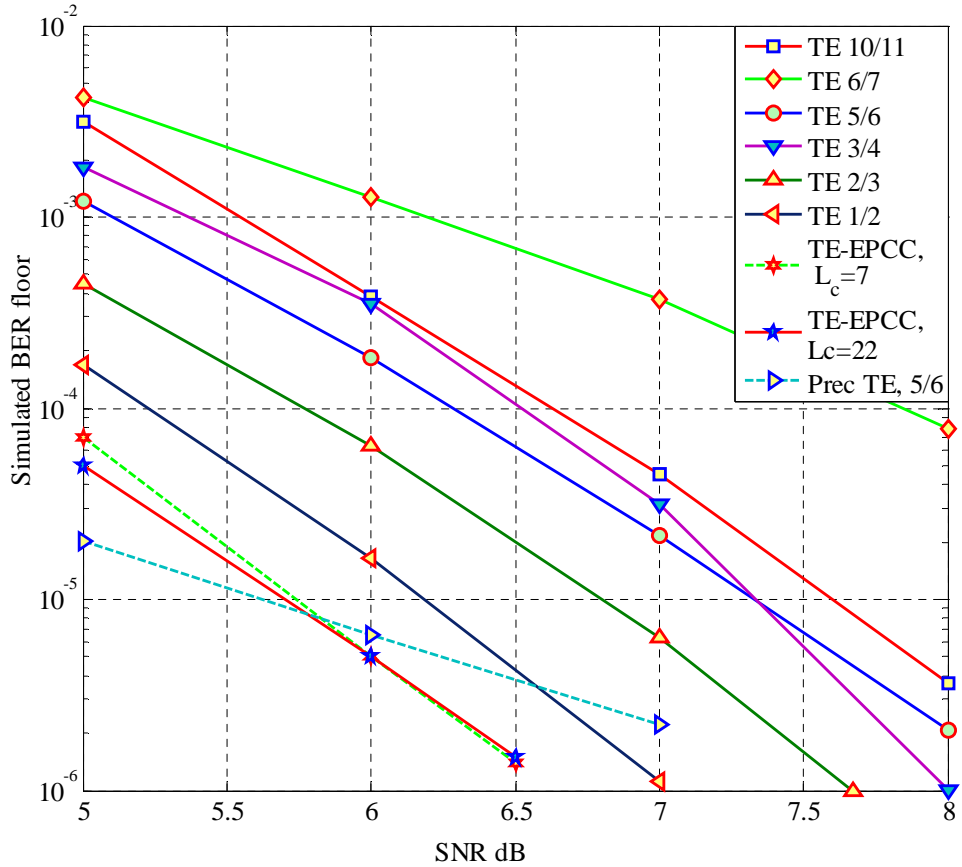}
\caption{Simulation of the TE-EPCC and the conventional TE for
various rates. }\label{diffR_sim}
\end{figure}

\begin{figure}[hbtp]
\centering
\includegraphics*[width= \figsz in]{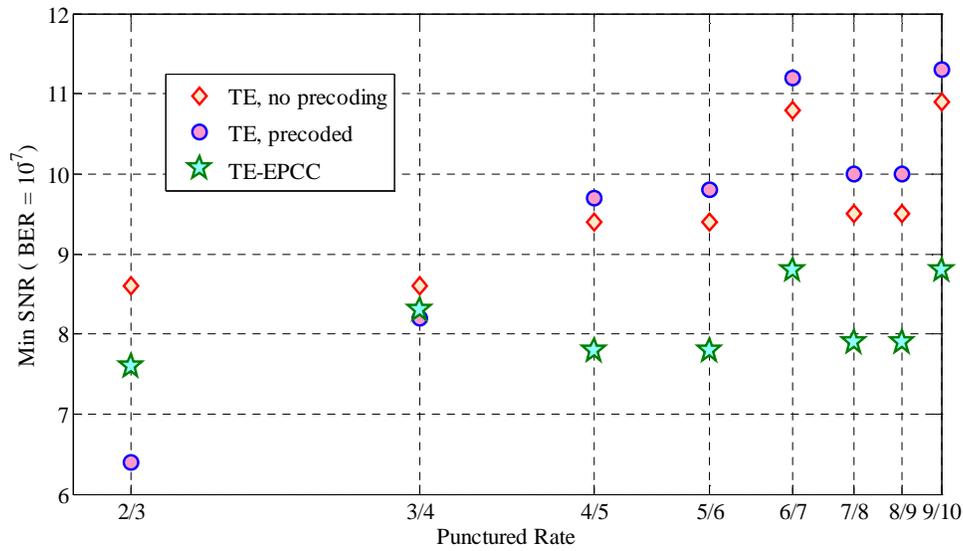}
\caption{Minimum SNR required to achieve $10^{-7}$ BER for various
TE systems as function of the outer $(7,5)$ RSCC punctured rate,
$N=1200$, and $\{m_c=3,L_c=1,d_c=10\}$ TE-EPCC.}\label{diffR}
\end{figure}

We also wish to study TE-EPCC advantage at various total system
rates and distributions of redundancy between the outer RSCC and
inner EPCC subcode. In Fig.~\ref{diffR_sim}, we compare the
simulated BER of the conventional non-precoded TE and the TE-EPCC,
for interleaver length $N=4312$, different punctured-rate RSCC
with connections $(7,5)$, and EPCC with $m_c=3$ and $d_c=10$. The
results show that the TE-EPCCs composed of either $L_c=7$
$(630,616)$ EPCC or $L_c=22$ $(210,199)$ EPCC concatenated to rate
$\frac{5}{6}$ TE, achieve the same BER in the error floor region.
Furthermore, they both outperform comparable rate conventional
TEs, with $L_c=22$ TE-EPCC furnishing a gain of $1.5$ dB with
respect to rate $\frac{3}{4}$ TE at BER=$10^{-6}$, and $L_c=7$
TE-EPCC delivering similar gain over rate $\frac{5}{6}$ TE.
Moreover, either TE-EPCCs deliver $1$ dB SNR gain over the
precoded TE of punctured-rate $\frac{5}{6}$. For a complete
investigation of a wide range of coding rates, we plot the minimum
SNR required to achieve a BER of $10^{-7}$ for punctured coding
rates from $\frac{2}{3}$ to $\frac{9}{10}$, comparing the
conventional non-precoded and precoded TE to the TE-EPCC. Such a
comparison is shown in Fig.~\ref{diffR} for interleaver length
$N=1200$, different punctured-rate RSCC with connections $(7,5)$,
and $L_c=1$ EPCC with $m_c=3$ and $d_c=10$. We conclude from the
results that the TE-EPCC delivers a uniform gain of $1.5$ dB for
puncturing rates above $\frac{3}{4}$. The abnormal peak in BER for
puncturing rate $\frac{6}{7}$ is due to the particular choice of
puncturing table. The reason why the precoded TE outperforms the
TE-EPCC for puncturing rates $\frac{2}{3}$ and $\frac{3}{4}$ can
be explained by examining $\mathbf{A}^o(2)$ for those puncturing
rates, where it was shown in~\cite{HakimPhD}, using a similar
approach to~\cite{Rowitch}, that the outer RSCC does not generate
Hamming weight $2$ errors for these low puncturing rates. Hence,
since the BER performance of precoded TE is dominated by such
errors in the floor region, its BER is significantly improved
surpassing the TE-EPCC at those rates. In summary, the precoded TE
is more effective when the minimum Hamming distance of the outer
code is larger than $2$. Hence, its less effective for high rate
simple punctured codes, where its hard to design punctured rates
of this property. Therefore, the TE-EPCC is more effective at high
code rates for which simple puncturing is utilized.


\section{Conclusion}

In this work, we have studied the BER of the serial concatenation
of EPCC and interleaved RSCC over ISI channels as an alternative
to a single RSCC with and without an inner precoder. To facilitate
the study of system performance for a wide range of coding rates,
interleaver sizes, and EPCC design parameters, we have derived an
approximate upper bound on the BER of the TE-EPCC that is easy to
evaluate and that scales well with system parameters. We have also
shown how EPCC enhances TE performance by reducing the frequency
of error words of low Euclidean distance, which dominate the BER
both in the waterfall and error floor regions. Numerical results,
calculated via the derived bound, indicate that the TE-EPCC
delivers substantial gain for short interleaver lengths compared
to the precoded and non-precoded TE, which makes it more
attractive than the conventional TE for hardware implementation.
Also, we have demonstrated that the TE-EPCC furnishes a uniform
gain of $1.5$ dB for puncturing rates above $0.75$, which makes it
suitable for high rate applications, such as magnetic and optical
recording applications, while the precoded TE is a better choice
for lower coding rates.

\appendix
\subsection{Simple BER bound expressions for $L_c=1$}
\subsubsection{Non-precoded TE} As discussed above for the $1-
\alpha D$ channel, the Euclidean distance of a compound error
event of multiplicity $m$ and $\gamma$ crossing branches is given
by:
\begin{eqnarray}\label{eqA1}
\nonumber d_E^2= 4 \alpha \gamma + (1-\alpha)^2 d + 2 \alpha m -
\mu \alpha^2.
\end{eqnarray}
where $d$ is the Hamming distance of the compound error. In order
to compare the channel SNR of different levels of ISI $\alpha$, we
make use of the noise variance normalization:
\begin{eqnarray}\label{eqA2}
\nonumber \tilde{\sigma}^2=\frac{1+\alpha^2}{2} \sigma^2
\end{eqnarray}
in which case the dicode channel has the base noise variance
$\sigma^2$. When the EPCC is turned off, the expression of the
TE's BER reduces to:
\begin{eqnarray}\label{eqA3}
\nonumber P_b & \leq & \frac{1}{K}\sum_{d_E=1}^{\infty}
Q\left(\frac{d_E}{\tilde{\sigma}}\right)\sum_{d=2}^{d_{T}}
 \frac{\mathbf{A}^o(d)\mathbf{\overline{A}}^i(d)}{{ N \choose d }} \\
\nonumber & & \sum_{\mu=0}^{1}\mathop {\sum_{m=1}^{d}}\limits_{
m:\ \gamma \geq 0, \gamma \in \mathbf{N}} { d-m \choose \gamma} \\
\nonumber & &
{\left(\frac{1}{2}\right)}^{d-m}{ N-d \choose m-\mu } { d-1 \choose m-1 }\\
\nonumber \gamma & = & \frac{d_E^2-2 \alpha m+ \mu \alpha^2 -
(1-\alpha)^2 d}{4\alpha}
\\
\end{eqnarray}
A good approximation of the Q function that is accurate for a wide
range of abscissa is borrowed from~\cite{Simon2003}, and is given
by:
\begin{eqnarray}\label{eqA4}
\nonumber Q\left(\frac{d_E}{\tilde{\sigma}}\right) \simeq
\frac{1}{12}e^{-\frac{d_E^2}{2\tilde{\sigma}^2}} +
\frac{1}{4}e^{-\frac{2d_E^2}{3\tilde{\sigma}^2}}.
\end{eqnarray}
Hence, $P_b$ in~(\ref{eqA3}) is composed of two terms as in:
\begin{eqnarray}\label{eqA5}
\nonumber P_b\left(\tilde{\sigma}\right) =
\frac{1}{12}\check{P_b}(\sqrt{2}\tilde{\sigma}) +
\frac{1}{4}\check{P_b}(\sqrt{\frac{3}{2}}\tilde{\sigma})
\end{eqnarray}
where
\begin{eqnarray}\label{eqA6}
\nonumber \check{P_b}(\tilde{\sigma}) & \leq &
\frac{1}{K}\sum_{d=2}^{\infty}
\mathbf{A}^o(d)\mathbf{\overline{A}}^i(d)  \\
\nonumber & & \sum_{\mu=0}^{1} \sum_{m=1}^{d}
\sum_{\gamma=0}^{d-m} { d-1 \choose m-1 } { d-m \choose \gamma }
{\left(\frac{1}{2}\right)}^{d-m} \\
\nonumber & & N^{m-\mu-d} \frac{d!}{(m-\mu)!}e^{-\frac{4 \alpha
\gamma + (1-\alpha)^2 d + 2 \alpha m -
\mu \alpha^2}{\tilde{\sigma}^2}}. \\
\end{eqnarray}
Evaluating the summation over $\gamma$ by utilizing the binomial
identity we obtain:
\begin{eqnarray}\label{eqA7}
\nonumber \check{P_b}(\tilde{\sigma}) & \leq &
\frac{1}{K}\sum_{d=2}^{\infty}
\frac{\mathbf{A}^o(d)\mathbf{\overline{A}}^i(d)}{(2N)^d}  \\
\nonumber & & \sum_{\mu=0}^{1} N^{-\mu} e^{-\frac{(1-\alpha)^2 d -
\mu
\alpha^2}{\tilde{\sigma}^2}}\left(1+e^{-\frac{4\alpha}{\tilde{\sigma}^2}}\right)^d
 \\ \nonumber & & \sum_{m=1}^{d}
\frac{d!}{(m-\mu)!} { d-1 \choose m-1
} \left(\frac{2Ne^{-\frac{2\alpha}{\tilde{\sigma}^2}}}{1+e^{-\frac{4\alpha}{\tilde{\sigma}^2}}}\right)^m. \\
\end{eqnarray}
After some algebraic manipulation, (\ref{eqA7}) simplifies to:

\begin{eqnarray}\label{eqA8}
\nonumber \check{P_b}(\tilde{\sigma}) & \leq &
\frac{1}{K}\sum_{d=2}^{\infty}
\frac{\mathbf{A}^o(d)\mathbf{\overline{A}}^i(d)}{(N)^d}  \\
\nonumber & & \sum_{\mu=0}^{1} N^{1-\mu} e^{-\frac{1+\alpha^2 ( d
- \mu)}{\tilde{\sigma}^2}}\frac{d!}{(1-\mu)!}
\left(\cosh\left(\frac{2\alpha}{\tilde{\sigma}^2}\right)\right)^{d-1}
\\ \nonumber & &
{}_1F_1\left(1-d; 2-\mu; -N\text{sech}\left(\frac{2\alpha}{\tilde{\sigma}^2}\right)\right)  \\
\end{eqnarray}
where ${}_1F_1$ is the confluent hypergeometric function of the
first kind~\cite{AeqB,Slater60}.

\subsubsection{TE-EPCC}

The BER of TE-EPCC, $P_b^{epcc}$, is expressed as the residual
error rate after subtracting the error rate component that is
correctable by EPCC, $P_b^{\mathcal{C}}$, from the conventional
non-precoded TE BER, $P_b$, as expressed in:

\begin{eqnarray}\label{eqA9}
\nonumber P_b^{epcc}\left(\tilde{\sigma}\right) =
P_b(\tilde{\sigma}) - P_b^{\mathcal{C}}(\tilde{\sigma}).
\end{eqnarray}
Similar to~(\ref{eqA4}), using the Q function approximation we
have
\begin{eqnarray}\label{eqA10}
\nonumber P_b^{\mathcal{C}}\left(\tilde{\sigma}\right) =
\frac{1}{12}\check{P_b}^{\mathcal{C}}(\sqrt{2}\tilde{\sigma}) +
\frac{1}{4}\check{P_b}^{\mathcal{C}}(\sqrt{\frac{3}{2}}\tilde{\sigma})
\end{eqnarray}
with
\begin{eqnarray}\label{eqA11}
\nonumber \check{P_b}^{\mathcal{C}}(\tilde{\sigma}) & \leq &
\frac{1}{K}\sum_{d=2}^{\infty}
\frac{\mathbf{A}^o(d)\mathbf{\overline{A}}^i(d)}{(2N)^d}\sum_{\mu=0}^{1}
N^{-\mu} e^{-\frac{(1-\alpha)^2 d -
\mu \alpha^2}{\tilde{\sigma}^2}}  \\
\nonumber & &  \sum_{m=1}^{\min(d,m_c)} \frac{d!}{(m-\mu)!} { d-1
\choose m-1
}\left(2Ne^{-\frac{2\alpha}{\tilde{\sigma}^2}}\right)^m \\
\end{eqnarray}
which can be expanded into two sum terms depending on the value of
$d$:
\begin{eqnarray}\label{eqA12}
\nonumber \check{P_b}^{\mathcal{C}}(\tilde{\sigma}) & \leq &
\frac{1}{K}\sum_{d=2}^{m_c}
\frac{\mathbf{A}^o(d)\mathbf{\overline{A}}^i(d)}{(2N)^d}\sum_{\mu=0}^{1}
N^{-\mu} e^{-\frac{(1-\alpha)^2 d -
\mu \alpha^2}{\tilde{\sigma}^2}}   \\
\nonumber & & \sum_{m=1}^{d} \frac{d!}{(m-\mu)!} { d-1 \choose m-1
}\left(2Ne^{-\frac{2\alpha}{\tilde{\sigma}^2}}\right)^m \\
\nonumber & & +\frac{1}{K}\sum_{d=m_c+1}^{\infty}
\frac{\mathbf{A}^o(d)\mathbf{\overline{A}}^i(d)}{(2N)^d}\sum_{\mu=0}^{1}
N^{-\mu} e^{-\frac{(1-\alpha)^2 d -
\mu \alpha^2}{\tilde{\sigma}^2}}  \\
\nonumber & &  \sum_{m=1}^{m_c} \frac{d!}{(m-\mu)!} { d-1 \choose
m-1
}\left(2Ne^{-\frac{2\alpha}{\tilde{\sigma}^2}}\right)^m. \\
\end{eqnarray}
After some algebraic manipulation and gathering of geometric
series terms we obtain the simplified expression:
\begin{eqnarray}\label{eqA13}
\nonumber \check{P_b}^{\mathcal{C}}(\tilde{\sigma}) & \leq &
\frac{1}{K}\sum_{d=2}^{\infty}
\frac{\mathbf{A}^o(d)\mathbf{\overline{A}}^i(d)}{(2N)^d}\sum_{\mu=0}^{1}
N^{-\mu} e^{-\frac{(1-\alpha)^2 d -
\mu \alpha^2}{\tilde{\sigma}^2}}  \\
\nonumber & & 2N(d!)e^{-\frac{2\alpha}{\tilde{\sigma}^2}}
{}_1F_1\left(1-d; 2-\mu; -2Ne^{-\frac{2\alpha}{\tilde{\sigma}^2}}\right)  \\
\nonumber & & -\frac{1}{K}\sum_{d=m_c+1}^{\infty}
\frac{\mathbf{A}^o(d)\mathbf{\overline{A}}^i(d)}{(2N)^d}\sum_{\mu=0}^{1}
N^{-\mu} e^{-\frac{(1-\alpha)^2 d -
\mu \alpha^2}{\tilde{\sigma}^2}}  \\
\nonumber & &  \frac{d!}{(m_c+1-\mu)!} { d-1 \choose m_c } (2Ne^{-\frac{2\alpha}{\tilde{\sigma}^2}})^{m_c+1}\\
\nonumber & & {}_2F_2\left(1,m_c+1-d; m_c+1, m_c+2-\mu; -2Ne^{-\frac{2\alpha}{\tilde{\sigma}^2}}\right)\\
\end{eqnarray}
where ${}_2F_2$ is the generalized hypergeometric
function~\cite{Slater66}.

\subsubsection{precoded TE} By examining the precoded trellis in
Fig.~\ref{precoded_Tr}, the squared Euclidean distance of a
compound error of Hamming distance $d$ and length $L$ for the
precoded channel $\frac{1-\alpha D}{1 \oplus D}$ is:

\begin{eqnarray}\label{eqA14}
\nonumber d_E^2= \lceil \frac{d}{2} \rceil +
\lfloor\frac{d}{2}\rfloor \alpha^2 + 4\alpha \gamma +
(1-\alpha)^2(L-d).
\end{eqnarray}
 Substituting this expression in the bound on BER of the precoded TE
 derived in~\cite{Oberg01}, and utilizing the approximation of the Q function once again, we
 obtain:
\begin{eqnarray}\label{eqA15}
\nonumber \check{P_b}(\tilde{\sigma}) & \leq &
\frac{1}{K}\sum_{d=2}^{N}
 \frac{\mathbf{A}^o(d)\mathbf{\overline{A}}^i(d)}{{ N \choose d }}  \\
\nonumber & & \sum_{L=d}^{N}\left(\frac{1}{2}\right)^{L-d} {
N-L+\lfloor\frac{d}{2}\rfloor \choose \lfloor\frac{d}{2}\rfloor }
{ L-1-\lceil \frac{d-1}{2} \rceil \choose
\lfloor\frac{d-1}{2}\rfloor }  \\
\nonumber & & e^{-\frac{(1-\alpha)^2 (L-d) + \lceil \frac{d}{2}
\rceil + \lfloor\frac{d}{2}\rfloor
\alpha^2}{\tilde{\sigma}^2}}\sum_{\gamma=0}^{L-d} { L-d \choose
\gamma } e^{-\frac{4 \alpha \gamma
}{\tilde{\sigma}^2}}. \\
\end{eqnarray}
Evaluating the summation over $\gamma$ by utilizing the binomial
identity, we obtain after some simplification:
\begin{eqnarray}\label{eqA16}
\nonumber \check{P_b}(\tilde{\sigma}) & \leq &
\frac{1}{K}\sum_{d=2}^{N}
 \frac{\mathbf{A}^o(d)\mathbf{\overline{A}}^i(d)}{{ N \choose d }}  e^{-\frac{\lceil \frac{d}{2} \rceil +
\lfloor\frac{d}{2}\rfloor \alpha^2}{\tilde{\sigma}^2}} \\
\nonumber & & \sum_{L=d}^{N} { N-L+\lfloor\frac{d}{2}\rfloor
\choose \lfloor\frac{d}{2}\rfloor } { L-1-\lceil \frac{d-1}{2}
\rceil \choose \lfloor\frac{d-1}{2}\rfloor
}\Psi(\alpha,\tilde{\sigma})^{L-d}\\
\nonumber \Psi(\alpha,\tilde{\sigma}) & = & e^{-\frac{1+\alpha^2}{\tilde{\sigma}^2}}\cosh\left(\frac{2\alpha}{\tilde{\sigma}^2} \right) \\
\end{eqnarray}
This can be simplified to a single sum over $d$ by the utility of
the generalized hypergeometric representation, which is given by:
\begin{eqnarray}\label{eqA17}
\nonumber \check{P_b}(\tilde{\sigma}) & \leq &
\frac{1}{K}\sum_{d=2}^{N}
 \frac{\mathbf{A}^o(d)\mathbf{\overline{A}}^i(d)}{{ N \choose d }} { N-\lceil\frac{d}{2}\rceil \choose \lfloor\frac{d}{2}\rfloor
} e^{-\frac{\lceil \frac{d}{2} \rceil +
\lfloor\frac{d}{2}\rfloor \alpha^2}{\tilde{\sigma}^2}} \\
\nonumber & & {}_3F_2\left(1, \lfloor \frac{d+1}{2} \rfloor,d-N;
\lceil\frac{d}{2}\rceil-N,1 ;
\Psi(\alpha,\tilde{\sigma})\right). \\
\end{eqnarray}
When $N \gg d$, we can use similar approximations to the ones used
in the derivation of the interleaver gain exponent, by which one
reaches a looser, albeit simpler, bound:
\begin{eqnarray}\label{eqA18}
\nonumber \check{P_b}(\tilde{\sigma}) & \leq &
\frac{1}{K}\sum_{d=2}^{N}
 \mathbf{A}^o(d)\mathbf{\overline{A}}^i(d) \frac{d!}{\lfloor\frac{d}{2}\rfloor!}N^{-\lceil \frac{d}{2} \rceil} e^{-\frac{\lceil \frac{d}{2} \rceil +
\lfloor\frac{d}{2}\rfloor \alpha^2}{\tilde{\sigma}^2}} \\
\nonumber & & {}_3F_2\left(1, \lfloor \frac{d+1}{2} \rfloor,d-N;
\lceil\frac{d}{2}\rceil-N,1 ;
\Psi(\alpha,\tilde{\sigma})\right). \\
\end{eqnarray}

\bibliographystyle{IEEEtran}
\bibliography{TEEPCC_ref}

\end{document}